# Energy Scenario Exploration with Modeling to Generate Alternatives (MGA)

J.F. DeCarolis[†], S. Babaee, B. Li, S. Kanungo

Department of Civil, Construction, and Environmental Engineering
North Carolina State University Campus Box 7908
Raleigh, NC, USA 27695

[†] Corresponding Author.
Phone: +1-919-515-0480; Fax: +1-919-515-7908; Email: jdecarolis@ncsu.edu

**Software Availability**

| | |
|---|---|
| Name of software | Tools for Energy Model Optimization and Analysis (Temoa) |
| Developers | Joseph DeCarolis, Sarat Sreepathi, Kevin Hunter, Binghui Li, Suyash Kanungo |
| Contact | jdecarolis@ncsu.edu |
| Year First Available | 2012 |
| Hardware required | A personal computer |
| Software Required | Microsoft Windows, Mac OSX, or Linux operating system. Python, Pyomo, GLPK, Graphviz, Matplotlib |
| Software Availability | Temoa source code can be accessed through the project website: http://temoaproject.org or directly through Github: https://github.com/TemoaProject/temoa |
| Cost | All software elements are open source and freely available. Temoa is offered under the GNU General Public License, version 2. |

**Abstract**

Energy system optimization models (ESOMs) should be used in an interactive way to uncover knife-edge solutions, explore alternative system configurations, and suggest different ways to achieve policy objectives under conditions of deep uncertainty. In this paper, we do so by employing an existing optimization technique called modeling to generate alternatives (MGA), which involves a change in the model structure in order to systematically explore the near-optimal decision space. The MGA capability is incorporated into Tools for Energy Model Optimization and Analysis (Temoa), an open source framework that also includes a technology rich, bottom up ESOM. In this analysis, Temoa is used to explore alternative energy futures in a simplified single region energy system that represents the U.S. electric sector and a portion of the light duty transport sector. Given the dataset limitations, we place greater emphasis on the methodological approach rather than specific results.

## 1. Introduction

Effective mitigation efforts that avoid or limit dangerous anthropogenic influence with the climate require fundamental changes in the way energy is supplied and demanded globally over the next half century. Because energy infrastructure is expensive and long-lived, a critical challenge is to develop robust planning and investment strategies that account for future uncertainty. Energy system optimization models (ESOMs) represent a key tool that can be used to probe the future decision space under different future scenarios (DeCarolis 2011; DeCarolis et al. 2012). Such models calculate an intertemporal partial equilibrium on energy markets by optimizing the energy system over time in order to minimize cost or maximize surplus. ESOMs generally have a national to global scope and are optimized over several future decades in order to see the system response to exogenous conditions such as new policy implementation, fuel price shifts, or technology innovation.

Given the expansive physical and temporal system boundaries involved, ESOM-based analyses are faced with conditions of deep uncertainty. Deep uncertainty reflects circumstances in which stakeholders do not know or cannot agree on (1) the choice of models to accurately capture key system interactions, (2) the probability distributions associated with key uncertain parameters, and (3) how to value the desirability of outcomes (Walker et al., 2013). Disagreement over the choice of models reflects structural uncertainty, whereby the relationship among key modeled and unmodeled factors is not fully known (Lempert et al., 2003). All ESOMs are radical simplifications of complex real world phenomena and no single model structure can completely capture it (DeCarolis, 2011). In addition to imperfect models, the future values or even distributions of key uncertain parameters used to populate the models are often highly uncertain. Furthermore, it is not clear how best to value future outcomes; for example, through the choice of intertemporal discount rate. The difficulty in applying subjective, valued-based judgement to find socially desirable – or even acceptable – solutions led Rittel and Webber (1973) and Liebman (1976) to characterize ill-defined public planning problems as "wicked."

Given such deep uncertainty about the future, singular model projections have little or no value and can often be misleading. The focus should be on producing model-based insights rather than "precise-looking" projections; the latter can distract and unduly influence the planning process with false precision (Huntington et al., 1982; Peace and Weyant, 2008). A common approach to model-based analysis that avoids the pitfalls associated with forecasting is scenario analysis, where each scenario corresponds to a storyline about how the future may unfold along with a set of exogenous assumptions consistent with the storyline that are used to drive the model. However, as Morgan and Keith (2008) point out, scenarios with detailed storylines can play into cognitive biases by appearing more plausible and probable than they are in reality. Another limitation of scenario analysis is that mutually exclusive and exhaustive subjective probabilities are often not assigned to scenarios, leaving decision makers with a disparate set of energy futures to ponder (Morgan and Keith, 2008; Kann and Weyant, 2000). Finally, traditional scenario analysis can be effective with small groups of clients whose concerns are well known to the scenario developers, but can fail to generate consensus in broad public debates that include divergent interests and values (Bryant and Lempert, 2010).

Kann and Weyant (2000) assert that "ideal results" from uncertainty analysis with ESOMs would include probability-weighted model outputs, optimal decisions that account for imperfect information, a measure of risk or dispersion in the outcome, and the value of information associated with key variables. Such output metrics help inoculate model-based analysis from both false precision and cognitive heuristics. However, an overarching framework is required that enables users to iterate models, produce results, and formulate high-level insights that can be applied within the decision making process. For example, Computer-Assisted Reasoning (CAR) is an approach to decision making under deep uncertainty that enables efficient model iteration and enhanced user ability to interrogate model results through computer visualization and search (Lempert, 2002).

By contrast, most ESOM-based analyses are published with insights summarized by the authors, and do not demonstrate how the models can be used in an iterative approach to generate insights and inform decisions. This paper is a step towards addressing this deficiency. The purpose of this paper is to illustrate how an ESOM can be used to explore alternative energy system designs under conditions of deep uncertainty using an optimization technique known as modeling to generate alternatives (MGA). By generating a sequence of near optimal solutions that are very different in decision space, MGA can produce alternatives for further evaluation by the analyst. While DeCarolis (2011) discussed the utility of MGA in an energy systems context, this paper represents the first published application of MGA to an ESOM.

To conduct the analysis, we use Tools for Energy Model Optimization and Analysis (Temoa), an open source, bottom-up energy system model (Hunter et al., 2013) along with a simplified input dataset constructed for this analysis. The dataset is focused on the U.S. electric and light duty transportation sectors, and can capture sector interactions through the deployment of plugin electric vehicles (PEVs) that require electricity for charging. Given the limited dataset used for this analysis, we place greater emphasis on the methodological approach rather than specific results. Our intention is to illustrate how an iterative approach to modeling using MGA can lead to insights that might not be realized through conventional scenario analysis.

## 2. An Electric and Transportation Sector Case Study

Together, the U.S electric and light duty transportation systems account for approximately 60% of national $CO_2$ emissions (U.S. EIA, 2015). Following the OPEC oil embargo, which led to the retirement of nearly all U.S. oil-fired power plants, the electric and transportation sectors have evolved more or less independently, with petroleum representing 0.7% of U.S. electricity supply, and 91% of light duty transportation (U.S. EIA, 2015). However, PEVs – including both plugin hybrid electric vehicles (PHEVs) and battery electric vehicles (BEVs) – have been rapidly deployed over the last 5 years and may lead to a significant coupling of the electric and transport sectors in the future. Given the threat of climate change, both sectors represent key targets for $CO_2$ emissions reductions. While there have been several Congressional bills that mandate a federal cap-and-trade program for greenhouses gases, none have been implemented (U.S. EPA, 2015). This analysis is focused on using MGA to explore different technology pathways to achieve a low carbon energy future. Prior to applying MGA, we ran three scenarios for comparative purposes: a base case scenario with no policy as well as moderate and aggressive climate policy scenarios. The moderate climate scenario includes a cap on $CO_2$ emissions that begins in 2025 and decreases linearly to 40% below 2015 values by 2050. The aggressive climate scenario also begins in 2025, but requires an 80% decrease below 2015 levels by 2050. These scenarios serve as a useful benchmark for the MGA runs. We then apply MGA to the moderate climate policy scenario in order to search for alternative, cost- and emissions-constrained solutions. Applying MGA in this way allows us to efficiently and systemically explore the model decision space. The resultant solutions can be used to characterize the tradeoff between system cost and emissions, and to identify alternative technology deployments that may be preferable to the original ones. While some MGA solutions may have higher cost, they may have appealing attributes to the planner if they capture unmodeled issues.

## 2.1 Model Description

We have developed Tools for Energy Model Optimization and Analysis (Temoa), a bottom-up, technology rich ESOM embedded within a larger framework for analysis. Temoa includes two key features that make it unique within the energy modeling community: (1) all source code and data are publicly archived online using a modern revision control system (TemoaProject, 2015), and (2) the model was designed to operate in a high performance computing environment in order to facilitate rigorous uncertainty analysis (Hunter et al., 2013). Temoa utilizes linear programming techniques to minimize the system-wide cost of energy supply by optimizing the deployment and utilization of energy technologies over a user-specified time horizon to meet end-use demands. The model is subject to a number of constraints that ensure proper system performance, including conservation of energy at the process and system-wide levels. In addition, users can impose additional constraints such as emissions bounds, minimum or maximum capacity and activity constraints, and growth rate limits. Model outputs by future time period include the optimal installed technology capacity and utilization, marginal energy prices, and emissions. Temoa, like many other bottom-up ESOMs, assumes rational decision making, with perfect information and perfect foresight, and simultaneously optimizes all decisions over the user-specified time horizon. Because the end-use demands remain fixed and are therefore unresponsive to price, Temoa represents a simplified partial equilibrium model.

## 2.2 Data Description

We developed a single region U.S. database compatible with Temoa that contains projected fuel prices, technology cost and performance estimates, and end-use demands. Here we provide a brief summary of key data elements relevant to this study.

The model time horizon is 2015 to 2050, with 5-year time periods. Each year within a given 5-year time period is assumed to have identical characteristics. Diurnal variation in renewable resource availability is represented by specifying four time segments (i.e., morning, mid-day, afternoon/evening, and night). For simplicity, we neglect seasonal variability in renewable energy supply and end use demands.

The high-level organization of the input database is provided in Fig. 1, which contains a simplified representation of the fuel supply, electric, and light duty vehicle (LDV) sectors. A complete view of the modeled energy system that includes a representation of all technologies and associated commodity flows is presented in Fig. A.1 in Appendix A. We are using the national U.S. TIMES dataset (NUSTD) as the main data source (Babaee, 2015). NUSTD is a single region, national-level input dataset developed for use with The Integrated MARKAL-EFOM System (TIMES) model generator (Energy Modeling, 2015). We have made a series of technology updates based on the Annual Energy Outlook (AEO) (U.S. EIA, 2015).

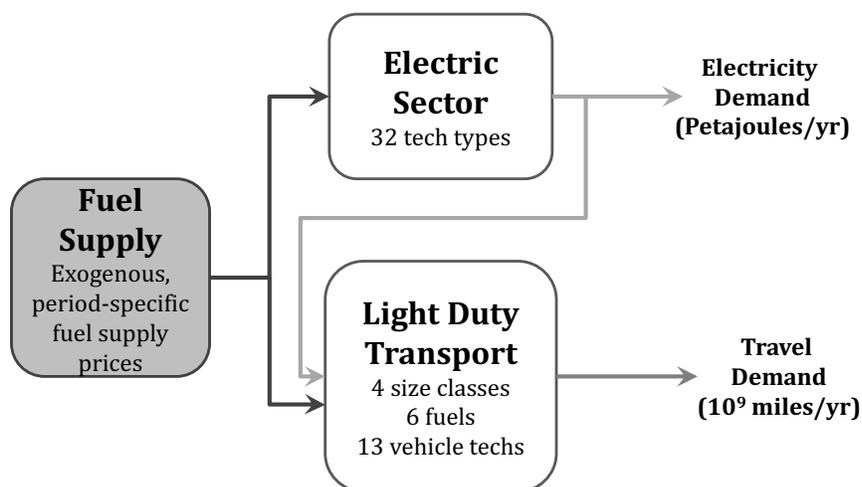

**Fig. 1.** Illustration of the modeled system, which includes an electric and light duty vehicle sector. Projected fuel prices are drawn from U.S. EIA (2012). Electricity prices are determined endogenously and therefore affect the cost-effectiveness of plug-in vehicles relative to other vehicle types. The model is driven by separate end use demands for electricity and travel distance.

Fuel supply is represented by a set of exogenously specified fuel price projections drawn from U.S. EIA (2012) with linear projections from 2040 to 2050. The electric sector contains 32 generation technologies and two carbon capture and storage (CCS) retrofit technologies to capture $CO_2$ emissions from new coal-fired steam and integrated gasification combined cycle (IGCC) power plants. The cost and performance data for the electric generators and the emission factors associated with the fuels consumed in the power plants are taken from NUSTD. The existing capacities and lifetime of the electric generators, shown in Table A.2 of Appendix A, are updated based on U.S. EIA (2015) and NREL (2010), respectively. We specify upper bounds on available renewable capacity based on AEO (2015). To further constrain electric technology performance within plausible limits, most new generating technologies are limited to an initial 30 GW of installed capacity in any post-2015 time period and an annual growth rate of 10% in subsequent time periods. These limits allow new technologies to reach a maximum of 500 GW installed capacity by 2050. Exceptions include new pulverized coal and nuclear, which were modeled with different limits. Given existing air quality regulations (EPA, 2013a; EPA, 2012) and the impending EPA $CO_2$ rules (EPA, 2014; EPA, 2013b), we limit initial new coal installation to 1 GW with a 10% growth rate, thereby constraining new pulverized coal to a maximum of 20 GW by 2050. Similarly, new nuclear was limited to 5 GW of installed capacity with a 10% growth rate, thereby constraining new nuclear to a maximum of 75 GW by 2050. While these assumptions limit coal and nuclear deployment, they allow for significant development beyond levels observed in the U.S. EIA (2015) scenarios. Tables

A.3 and A.4 in Appendix A contain the upper bounds on capacity values, growth rates, and growth rate seeds for new power plants in each time period. Aggregate electricity demand for the end-use sectors (commercial, industrial, residential, and heavy duty transportation) are exogenously specified in the model and drawn from U.S. EIA (2015), as shown in Table A.5 in Appendix A.

The light duty transportation sector includes 48 light duty vehicle technologies, which consist of 4 vehicle size classes, 6 fuel types, and 13 vehicle types. The 4 modeled LDV size classes are mini-compact, compact, full, and small SUV. Since the focus of the case study is on the potential link between the electric and transportation sectors through PEV deployment, for simplicity we do not model the larger vehicle sizes where U.S. EIA (2015) assumes PEV deployment will be negligible through 2040.

The 13 modeled vehicle-fuel types are conventional gasoline blended with 10% ethanol (E10), conventional gasoline blended with 85% ethanol (E85), diesel, compressed natural gas (CNG), hydrogen fuel cell vehicles, E10 hybrids, E85 hybrids, diesel hybrids, E10 plug-in hybrids with an all-electric range of 20 km (PHEV20), plug-in hybrids with an all-electric range of 60 km (PHEV60), E85-PHEV20, E85-PHEV60, and battery electric vehicles with an all-electric range of 160 km (BEV160). Vehicle lifetimes, capital costs, fuel economy, and $CO_2$ emission coefficients are obtained from NUSTD (Energy Modeling, 2015). The model accounts for both vehicle combustion emissions as well as upstream emissions associated with resource extraction. Table A.6 presents the assumed existing stock of LDVs by size class and fuel type, which is drawn from the 2010 existing capacity in NUSTD and linearly retired over the assumed 15 year lifetime of the vehicles (Energy Modeling, 2015). The total projected demand for vehicle miles associated with the four modeled LDV size classes, shown in Table A.7, is based on the projected annual growth rate of 1.3% for light duty transportation demand (U.S. EIA, 2012). The fixed percentage share of each vehicle size class in the LDV sector is presented in Table A.8 and is based on Shay et al. (2006).

A 5% social discount rate is used to discount future costs to the base year (2015). All alternative vehicle technologies (excluding gasoline and E85) have a 10% technology-specific discount rate (i.e., hurdle rate), which replaces the 5% discount rate when amortizing capital cost over the vehicle lifetime. Hurdle rates are used to adjust the amortized cost of alternative fuel vehicles relative to conventional gasoline vehicles in order to partially capture non-market factors that may affect their deployment, such as range anxiety or general aversion to new vehicle technology. While survey-based studies have estimated hurdle rates for alternative vehicle purchases in the range of 20-50 (Peterson and Michalek, 2013; Mau et al., 2008; Horne et al., 2005), our previous work indicates that applying even a 15% hurdle rate results in zero PEV deployment across a wide range of scenarios (Babaee et al., 2014). As a result, we assume that consumers make decisions based largely on vehicle cost-effectiveness.

We made several assumptions regarding the U.S. energy market and policy, which apply universally to the base case, the two $CO_2$ cap scenarios, and all the MGA runs. All scenarios include U.S. EIA (2012) reference case projections of oil and natural gas prices. To increase the viability of PEVs relative to other alternatives, we assume the attainment of program goals set forth by the DOE's Office of Energy Efficiency and Renewable Energy, which assumes a battery cost of 135 $/kWh in 2035 (U.S. EIA, 2011). In the MGA runs, we constrain 2015 technology activity to that observed in the base case to avoid the model optimizing a historical year.

Because we did not explicitly include the implementation of the Mercury and Air Toxics Standards (MATS) and the Cross-State Air Pollution Rule (CSAPR), which limit $SO_2$ and $NO_x$ emissions from the electric sector (EPA, 2012; EPA, 2013a), we apply an upper bound constraint, listed in Table A.9, on electricity generation from existing coal-fired power plant based on the reference case AEO projection to 2040 (AEO, 2015). Table A.10 represents the upper bound on ethanol availability from 2015-2025, which is obtained from the Renewable Fuel Standard (EPA, 2013c) and held constant from 2030 to 2050.

All data was entered and stored in SQLite, a relational database management system. Both the raw SQL file and the SQLite database are publicly available through our GitHub repository (TemoaProject, 2015).

## 3. Methods

The three modeled scenarios – a base case and two $CO_2$-constrained cases – provide some indication of which technologies can be most cost-effectively deployed to meet demand with and without an emissions limit. However, drawing insight from three scenarios is likely to be misleading because it neglects the deep uncertainties related to future energy system development. In order to systemically explore the model decision space while accounting for future uncertainty, we have implemented MGA within Temoa. Rather than generate different scenarios based on differing exogenous assumptions, the MGA algorithm changes the underlying structure of the mathematical model to search the feasible, near-optimal region of the solution space for alternative solutions that are very different in decision space. By changing the model structure, MGA finds solutions that may perform better when unmodeled objectives or constraints are considered exogenously. While parametric sensitivity analysis or Monte Carlo simulation could help identify different system configurations, it does not account for structural uncertainty in the model. As noted in DeCarolis (2011), the MGA results have an equally valid interpretation as perturbations of the objective function coefficients, and in this way also account for parametric uncertainty.

### 3.1 Hop-Skip-Jump MGA

While MGA can be formulated in a variety of ways, we begin by outlining the Hop-Skip-Jump (HSJ) MGA method described in Brill et al. (1982). The steps associated with HSJ MGA are as follows: (1) obtain an initial optimal solution by any method; (2) add a user-specified amount of slack to the value of the objective function(s); (3) encode the adjusted objective function value(s) as an additional upper bound constraint(s); (4) formulate a new objective function that minimizes the sum of decision variables that appeared in the previous solutions; (5) iterate the re-formulated optimization; and (6) terminate the MGA procedure when no significant changes to decision variables are observed in the solutions. Brill et al. (1982) formulate the revised MGA model described in Steps 3-4 above as follows:

Minimize:

$$p = \sum_{k \in K} x_k$$

Subject to:

$$f_j(\vec{x}) \leq T_j \quad \forall j$$
$$\vec{x} \in X$$

where $K$ represents the set of indices of decision variables with nonzero values in the previous solutions, $f_j(\vec{x})$ is the $j$th objective function in the original formulation, $T_j$ is the target value, including slack, specified for the $j$th modeled objective, and $X$ is the set of feasible solution vectors. $\vec{x} \in X$ implies that constraints in the original problem formulation also apply in the MGA formulation. The formulation above assumes that the objective function coefficients are unity, such that decision variables with non-zero values are given an equal weight of +1 in the first MGA iteration. In subsequent MGA iterations, each objective function coefficient can be incremented by +1 each time the associated decision variable takes on a positive value in a previous solution. Note that Steps 2-3 require encoding the original objective function as a constraint in the MGA formulation, with the scalar righthand side ($T_j$) set to the optimal objective function value plus some added slack. In this analysis, we test different slack values that represent a percentage increase in the base case objective function value. The Temoa objective function value represents the present cost of energy supply over the model time horizon. In this way, the MGA objective function attempts to minimize the decision variables that appeared in all previous solutions while constraining the system within a cost limit determined by the prescribed slack. The result is a sequence of model solutions, where each new solution considers prior solutions and is as far away from them as possible in decision space. While MGA could be used to return an arbitrarily large set of alternative system configurations, the intent is to provide a limited number of solutions that can be carefully evaluated by a human analyst.

## 3.2 MGA Customization

While MGA is a simple algorithm that can be applied to any optimization model, it should be tested and customized to better suit the specific modeling context in which it is applied. In applying MGA to an ESOM, we first had to decide which decision variables to consider within the MGA objective function. We considered two basic options: total activity by technology (i.e., total energy output over the model time horizon) and cumulative installed capacity per technology (i.e., maximum capacity over the model time horizon). While the MGA algorithm could be used to maximize the differences in installed capacity, all capacity may not necessarily be used to meet the end-use demands. We decided that total activity was a more accurate measure of a given technology's contribution to meeting demands within the energy system. To do this, we added a new derived variable to the model, V_ActivityByTech, which sums each technology's total output over the user-specified model time horizon. The updated source code for Temoa, including the MGA implementation described here, are publicly accessible through our Temoa Github site (TemoaProject, 2015).

Rather than consider all technology activity in the MGA objective function, we only consider the electric generating and vehicle technologies and ignore upstream processes related to fuel supply, fuel blending, and emissions accounting. This approach considers all production technologies on an equal basis and does not unfairly penalize technologies such as coal, which are linked to a larger number of upstream, accounting-related processes.

We test two MGA objective function formulations based on Brill et al. (1982). First, we account for the cumulative effect of deploying the same technology over multiple model iterations by incrementing its objective function weight by +1 after each model iteration with positive activity. For example, if coal produces 50,000 PJ in the base case, it receives a weight of +1 in the objective function associated with the first MGA iteration. If the resultant coal activity is reduced to 20,000 PJ in the first MGA iteration, its objective function weight in the second MGA iteration becomes +2. In the subsequent analysis, we refer to this as the 'integer' method.

Second, we tested a weighting scheme that uses normalized technology activity by sector as the weight in the MGA objective function. Again using coal as an example, suppose in the base case that total electricity generation over the entire model time horizon is 120,000 PJ, of which coal produces 50,000 PJ. The objective function weight for coal then becomes 0.42 in the first MGA iteration, which represents its fractional contribution to electricity supply in the base case. Further supposing that coal's fractional contribution is 0.10 in the first MGA iteration, its objective function weight becomes 0.52 in the second MGA iteration. Because each modeled sector may track technology activity in different units, this activity normalization must take place by sector. This second modification builds on the first by not only accounting for the cumulative effect of each technology's activity across different MGA iterations, but also accounts for the relative role of each technology within each sector. In the following analysis, we refer to this second MGA objective function weighting method as the 'normalized sector' method.

We test the relative performance of both the integer and normalized sector MGA weighting methods at different slack values. Adding slack to the objective function value gives the model space to select costlier alternative technology configurations compared to the original cost minimal solution. The higher cost associated with the MGA solutions can have two interpretations: (1) they account for parametric uncertainty related to the cost of energy technology deployment, and (2) they account for complex, unmodeled issues that suggest more expensive optimal solutions.

## 4. Results
The results are broken into several subsections. Section 4.1 presents results from the base and $CO_2$ cap scenarios, Section 4.2 presents comparative results from the two MGA weighting schemes, and Section 4.3 presents in-depth MGA results.

### 4.1 Results from the base and $CO_2$ cap scenarios
The base case results are presented in Fig. 2. Electricity production from coal remains relatively constant, while light water nuclear reactors and wind gain market share to make up for losses in gas turbine activity as natural gas prices increase toward mid-century. In the LDV sector, conventional gasoline vehicles remain dominant with an increased share of E85 vehicles in 2050 relative to 2015. No deployment of PEVs is present in the base case.

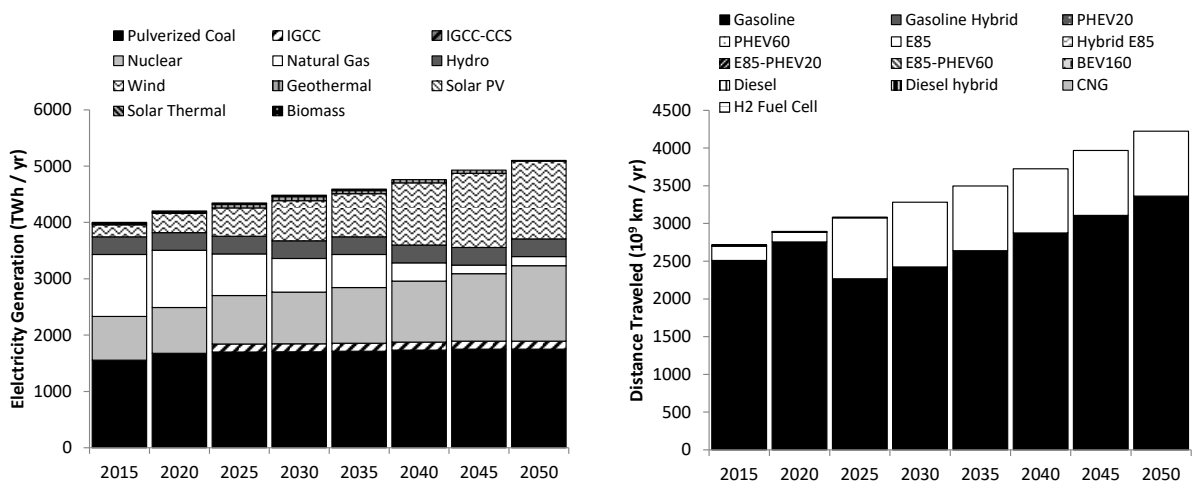

**Fig. 2.** Base case results from the electric sector (left) and light duty transportation sector (right).

Fig. 3 presents results from the CO₂ constrained scenarios. Both CO₂ caps are largely met through changes in the electric sector. In the 40% reduction scenario, the decline in coal generation is offset by increases in wind, and in 2050, integrated gasification combined cycle with carbon capture and sequestration (IGCC-CCS). Minimal changes occur in the LDV sector. In the 80% cap scenario, increasing shares of natural gas turbines and IGCC-CCS completely displace conventional coal by 2050. The LDV sector is still dominated by conventional gasoline vehicles, but battery electric vehicles (BEV160) reach a 21% market share by 2050.

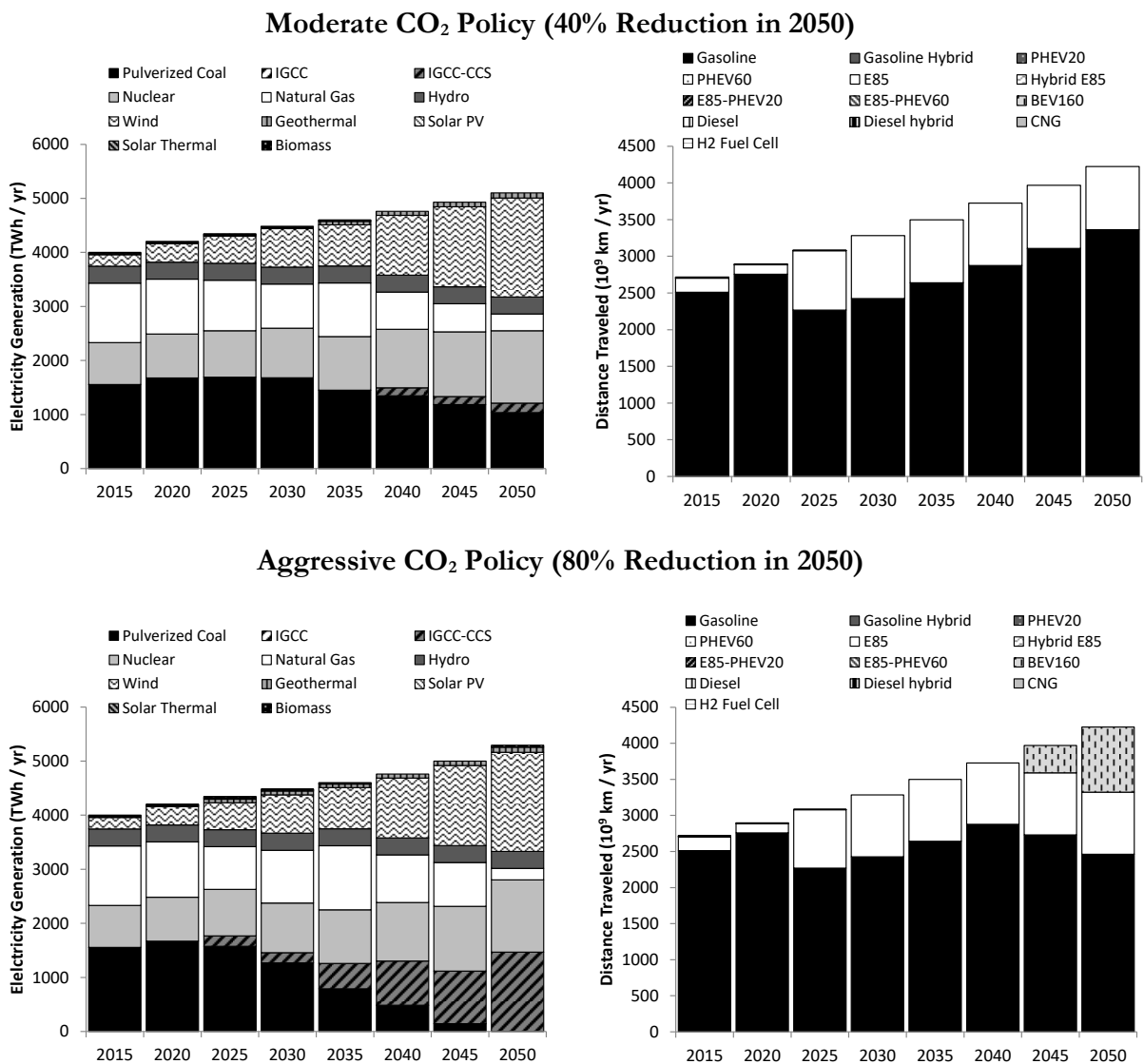

**Fig. 3.** Electric and LDV sector results from the 40% cap (top) and 80% cap (bottom) scenarios. The model largely utilizes a combination IGCC-CCS, wind, natural gas turbines, and battery electric vehicles (BEVs) to achieve the required emissions reductions.

## 4.2 MGA weighting scheme test

While the capped emissions scenarios appear plausible, it is impossible to ascertain the robustness of the results without further model introspection. Given the linear nature of the model, it is possible that these solutions sit on a knife-edge: perhaps small changes to key input parameters would reveal divergent new solutions. MGA is applied to look for alternative solutions that are very different in decision space but have a cost similar to the base and $CO_2$ capped solutions.

The 40% and 80% $CO_2$ reduction scenarios increase the present cost of energy supply over the model time horizon by approximately 0.4% and 1.9%, respectively, compared to the base case. We use these differences in costs to calibrate the slack in the MGA runs. Four sets of MGA runs were conducted with slack representing increases of 1%, 2%, 5%, and 10% over the base case present cost of energy supply. Given that we are using a highly simplified model to represent a complex system, we assume that even the 10% slack value is within the overall cost uncertainty of the model. At each of the four slack values, four MGA iterations are conducted to produce a total of 16 model runs for further examination. We apply MGA to the moderate climate policy scenario since it is only 0.4% more expensive than the base case and guarantees at least a 40% drop in 2050 $CO_2$ emissions relative to 2015. Such an approach can help characterize the available technology options in this cost- and emissions-constrained system given uncertainty related to both parameter values and model structure.

We need to choose metrics to compare the relative performance of the integer and normalized sector MGA formulations. Because we are searching for alternative technology configurations that can achieve low $CO_2$ emissions, we examine two indicators. First, we calculate the total number of unique energy technologies that are utilized in both the electric and light duty transport sectors across a set of 4 sequential MGA iterations at each slack value (Fig. 2). Calculating the number of deployed technologies provides a measure of the uniqueness associated with both MGA variants. Second, while $CO_2$ emissions are constrained to achieve a 40% reduction by 2050, we are interested in solutions that use the cost slack to achieve higher emissions reductions through different technology configurations.

Figure 4 indicates that the integer weighting method produces a larger number of deployed technologies compared to the normalized sector method. However, when using the integer method, the $CO_2$ cap is always binding such that 2050 emissions are 60% of 2015 levels in 2050. By contrast, the normalized sector method produces 2050 emissions levels that range from 16% to 60% of the 2015 emissions level. Although the integer method resulted in the deployment of more technologies, we chose to examine the normalized sector results in more detail because it delivered scenarios that resulted in emissions well below the specified cap. Thus the following results utilize the normalized sector method. For reference, the complete set of comparative results at all slack values is provided in Appendix B.

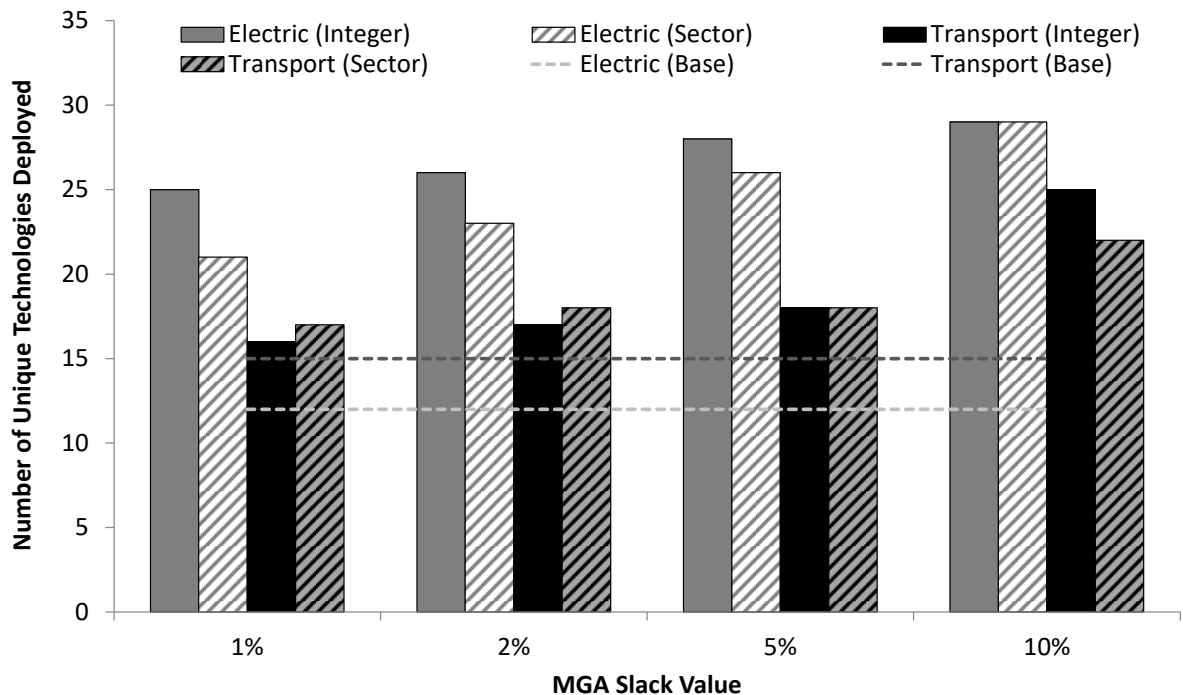

**Fig. 4.** Number of unique technologies utilized by sector, MGA weighting scheme, and assumed MGA slack value. For reference, the number of technologies utilized by sector in the moderate $CO_2$ cap scenario (without MGA) is included as a dotted line. Overall, the electric sector contains 34 technologies and the sector transport contains 48 light duty vehicle types.

### 4.3 In-depth MGA results

We wish to explore in more detail the MGA results associated with the normalized sector method. Trying to examine 16 sets of stacked bar plots for both the electric and transport sectors, as shown in Figs. 2 and 3, can be an overwhelming amount of information for an analyst to effectively evaluate. Instead, we have created a summary tableau (Fig. 5) that allows an analyst to survey the high level results and select scenarios of interest for further evaluation. While there are a number of ways that one could configure such a tableau, it should be adapted to display the model results that most directly address the issues and concerns at hand. Fig. 5 focuses on long-term results in 2050, including the total 2050 $CO_2$ emissions normalized by the 2015 base case $CO_2$ emissions, which enables direct comparison with the 40% and 80% emissions reductions required in the cap scenarios. Fig. 5 also includes the 2050 market share of various classes of technology by sector, which provides a high level view regarding which technologies make the largest contribution to meeting demand.

 

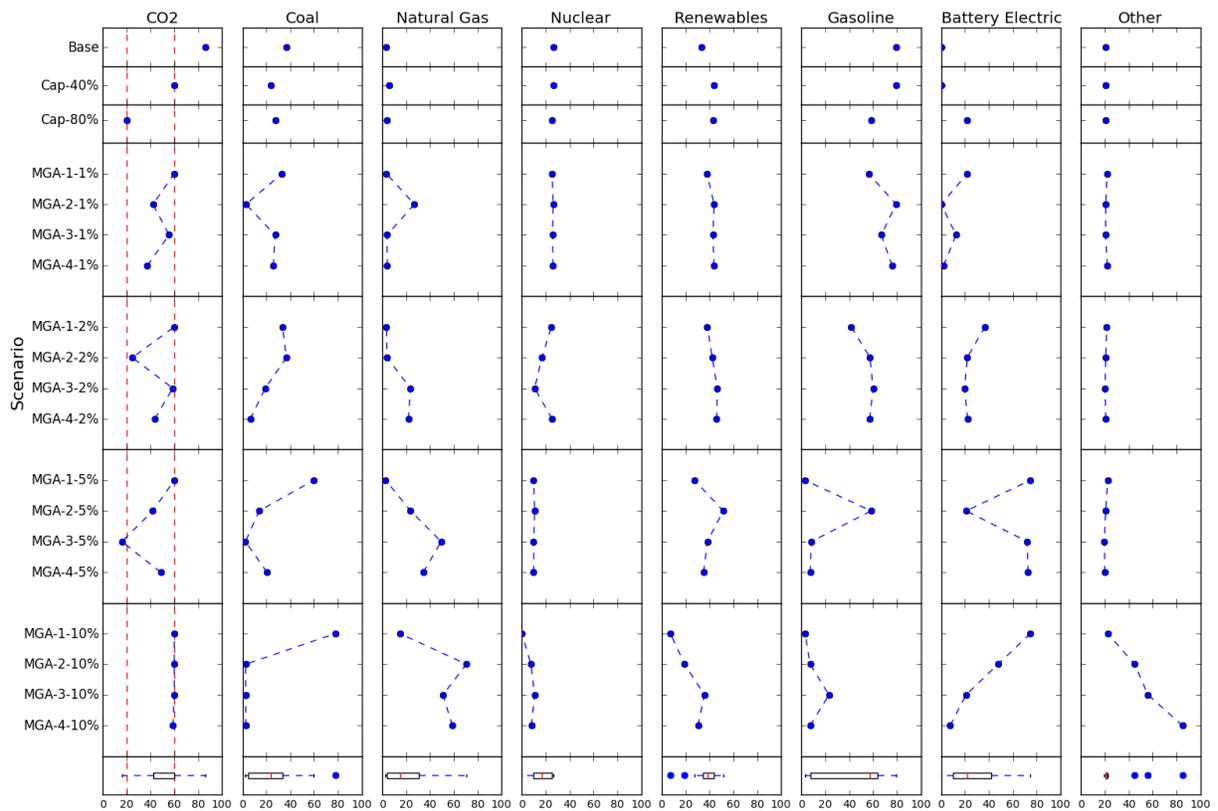

**Fig. 5.** Summary tableau for the base, $CO_2$ cap, and MGA runs. The vertical axis ticks indicate the model run; MGA runs are identified by their iteration number and slack value. For example, 'MGA-2-1%' represents MGA Iteration 2 with a 1% slack value. The leftmost column represents the 2050 $CO_2$ emissions expressed as a share of 2015 base emissions (the red dotted lines indicate the 2050 targets under the $CO_2$ cap scenarios); the next four columns represent the 2050 market share of each technology class within the electric sector. The three rightmost columns represent the 2050 market share of different classes of vehicle technology. Boxplots summarizing the distribution of points across all model runs and each technology class are included at the bottom.

Fig. 5 indicates that there are many possible low carbon technology configurations in the electric sector if we allow a modest increase in cost. Allowing a 1-2% increase in the present cost of energy supply, consistent with the cost of the $CO_2$ cap scenarios, can produce emissions reductions ranging from 40-75% in 2050 relative to 2015. For example, the fourth MGA iteration at 1% slack (MGA-4-1%) achieves a 63% reduction in 2050 $CO_2$ emissions, but only adds 0.6% more to the present cost compared the moderate $CO_2$ cap scenario shown in Figure 3. Those reductions are achieved largely with renewables (mainly wind), IGCC-CCS, and nuclear in the electric sector. With 1-2% slack, gasoline is still dominant in the LDV sector, with modest contributions from BEVs toward mid-century.

More detailed insights can be obtained by selecting specific model runs from Fig. 5 and examining the more detailed results in Appendix B. For example, MGA-3-2% represents a balanced approach that utilizes IGCC-CCS, nuclear, natural gas, renewables, and BEVs to achieve a 41% reduction in 2050 emissions. MGA-3-5% relies heavily on natural gas and renewables (biomass and wind) in the electric sector and BEVs to achieve an 84% reduction in $CO_2$ emissions. By contrast, MGA-4-10% utilizes a nearly 60% 2050 market share of natural gas in the electric sector as well as diesel, PHEV20, and BEVs in the LDV sector to achieve the required 40% reduction in 2050 emissions. Fig. 6 illustrates the variation in results: the top panel represents MGA-3-2% and the bottom panel represents MGA-4-10%. The complete set of results is presented in Appendix B.

### MGA Iteration 3 with 2% Slack (MGA-3-2%)

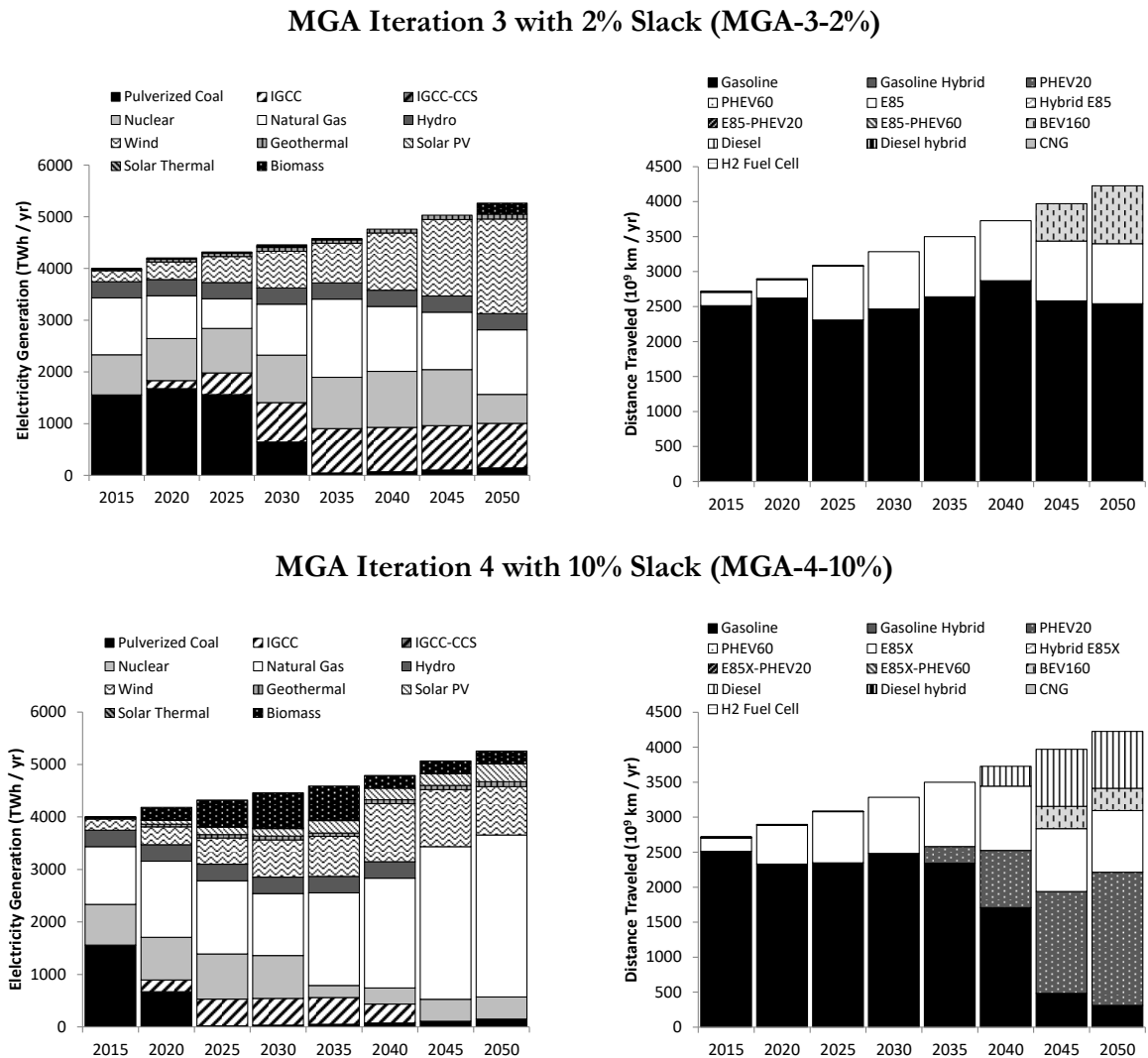

### MGA Iteration 4 with 10% Slack (MGA-4-10%)

**Fig. 6.** Select MGA results with an allowed 2% (top) and 10% (bottom) slack, which represents the increase in the present cost of energy supply over the model time horizon relative to the base case. The results illustrate the system design options under different cost constraints.

As the slack value increases, the variability in deployment by technology type also increases since higher slack values give the model more space under the cost constraint to deploy more expensive technologies. However, there are still many technologies that play little or no role across the 16 scenarios shown above. For example, solar photovoltaics (PV) do not obtain a significant market share and there is little deployment of alternative vehicles (other than BEVs) until a 10% slack value is employed.

As analysts, we may wonder if there is another way to produce a new set of alternative solutions. Another critical parameter in the model – and all ESOMs – is the choice of discount rate. The discount rate represents a value-related uncertainty because it determines the rate at which future costs should be valued relative to the same cost today. In Temoa, the discount rate is used to amortize the cost of capital investments over the technology lifetime and discount future costs back to the present. As such it affects the present cost of every technology modeled within the energy system and has the potential to produce significant shifts in technology adoption. Higher discount rates may shift deployment toward more capital intensive technologies, as those future investments will be more heavily discounted.

Thus far, a 5% social discount rate has been used. This value is a common choice in many analyses because it approximates the historical growth rate of the U.S economy (prior to the 2007 recession). While discounting future costs at roughly the growth rate of the economy is meant to represent a societal perspective, justification exists for the use of both lower and higher rates. To test the effect of discount rate on model solutions, we chose two bounding values: 0.1% and 10. We conduct this test at the intermediate 2% MGA slack value used in the preceding analysis.

A summary tableau of the discount rate tests are shown in Fig. 7. The non-MGA $CO_2$ cap scenario results at different discount rates are nearly identical, indicating that the change does not fundamentally alter the economic tradeoff between competing technologies. However, the MGA results at different discount rates indicate significant differences in technology deployment. The combination of a high discount rate and 2% MGA slack enable the model to deploy a wider suite of energy technologies, particularly in the LDV sector. Technology deployment with a 10% discount rate and 2% slack are qualitatively similar to the deployment pattern with a 5% discount rate and 5% slack. Higher discount rates can produce similar effects to higher MGA slack values by enabling the deployment of more capital intensive energy technologies, particularly near the end of the time horizon when present costs are relatively low. For example, higher deployments of capital intensive coal- and biomass-based IGCC as well as BEVs are possible in 2050 with 2% slack and a 10% discount rate compared to 2% slack and a 5% discount rate. Detailed results associated with discount rates of 0.1% and 10% are presented in Appendix C and can be compared to the results in Appendix B with a 5% discount rate.

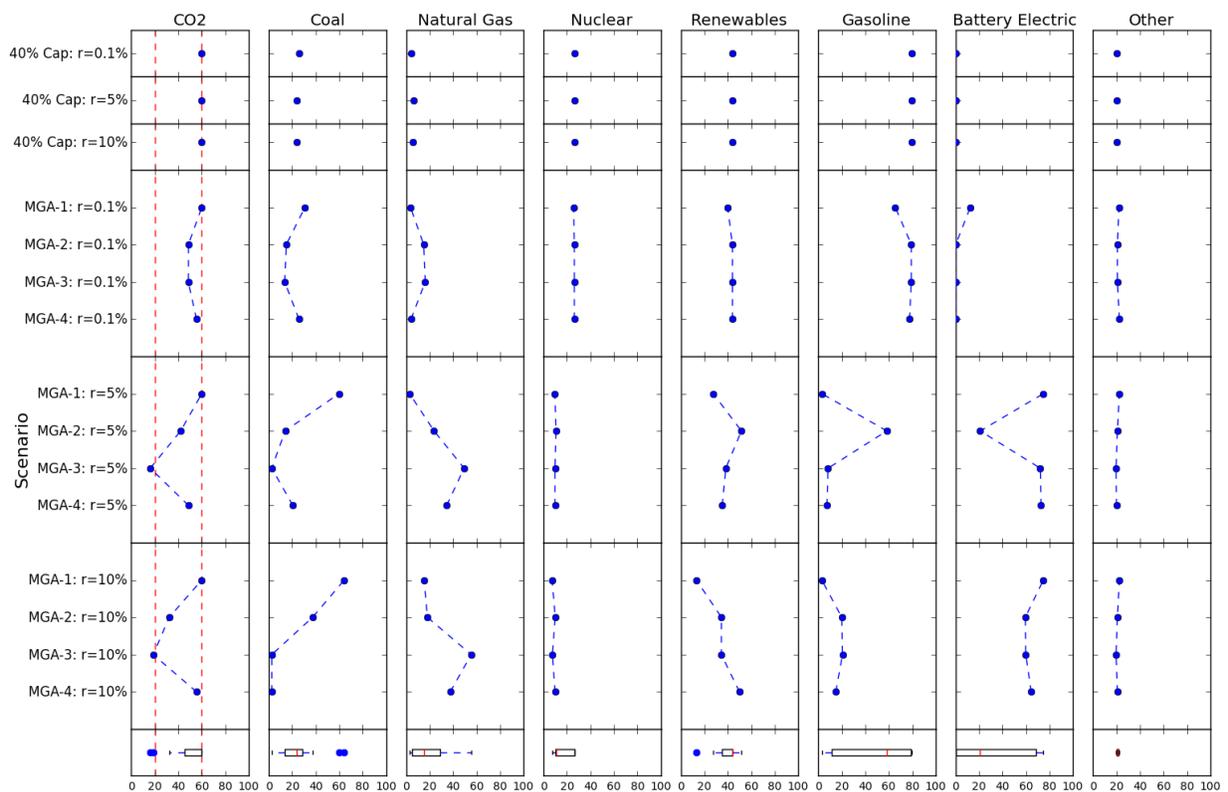

**Fig. 7.** Summary results from MGA tests in which discount rates of 0.1% and 10% were tested at a constant slack value of 2%. Results using the default 5% discount rate (also shown in Fig. 5) are included for comparison. $CO_2$ emissions represent 2050 emissions expressed as a fraction of the 2015 base year value (the red dotted lines indicate the 2050 targets under the $CO_2$ cap scenarios); the technology deployment columns represent the 2050 market share by sector and follow the same order as in Fig. 5.

Finally, suppose that as analysts, we are disappointed that the discount rate test did nothing to improve the prospects for solar PV, which only appeared in two model runs at an MGA slack value of 10% and discount rate of 5%, reaching a market share of 9% in 2050. We would like to quantify the contribution that solar PV can make when the slack is only 2%. Furthermore, we wish to determine whether PEVs can be deployed along with solar PV. In a final MGA test, we modify the sector normalized weighting method such that solar PV and all PEVs get a weight of 0 in a single MGA iteration. This approach will enable the model to meet demand using solar PV and PEVs without increasing the value of the MGA objective function. Note that the same approach could be taken to explore the maximum uptake of any single technology or set of technologies. The results are shown below in Fig. 8.

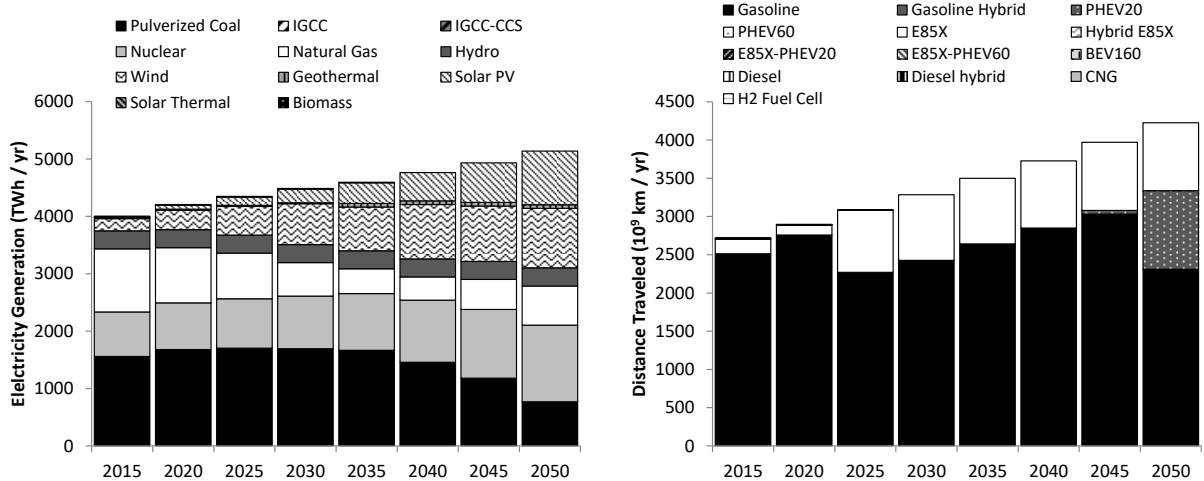

**Fig. 8.** MGA result with 2% slack and a 5% discount rate when the MGA objective function is set to maximize the deployment of solar PV and plugin vehicles.

As shown in Fig. 8, solar PV can reach a market share of 18% with a 2% MGA slack. The capacity results indicate that the annual 10% growth rate constraint on solar PV is binding, enabling it to reach an installed capacity of approximately 500 GW by 2050. While PEVs had zero weight in the MGA objective function, the run resulted in only a modest deployment of PHEV20 in 2050 because much of the cost slack was taken up through solar deployment. Further tests could be performed to seek balance between solar PV deployment and PEV deployment, if desired by the analyst. Another next step might be to revisit the 10% hurdle rate placed on alternative vehicles and see if reducing or eliminating it produces a significant effect on deployment. Such tests can be performed in an iterative fashion to probe the decision space and address specific questions posed by the analyst.

## 5. Discussion

MGA represents a simple method for systematically exploring the decision space of an energy system model. In this analysis, we produce alternative energy futures that help characterize system design options under a cost constraint. We also demonstrate how varying two scalar parameters – the MGA slack value and discount rate – can produce a diverse set of energy futures.

The MGA-based model results indicate that many technologies beyond those deployed in the base and CO$_2$ cap scenarios could play a significant role in a future energy system. For example, the MGA results include significant deployments of IGCC, biomass, and wind in the electric sector as well as BEVs in the LDV sector. Within the LDV sector, the model prefers to deploy BEVs as the alternative to gasoline and E85, and only deploys more expensive alternative vehicle technologies when the 10% MGA slack is available. More generally, the model tends to manipulate the electricity generation mix more readily than the vehicle mix because it can produce larger overall changes in technology activity at a smaller cost penalty. We also find that increasing the discount rate and the MGA slack value produce similar effects by enabling the deployment of more capital intensive technologies (e.g., nuclear and biomass-IGCC) in later model time periods. These results represent alternative futures that can either be dismissed quickly as implausible or present an intriguing option that warrants further investigation. An advantage of MGA is that the scenarios are generated by a computer algorithm and therefore lack the background detail that can lead to misleading, cognitively compelling storylines.

We note that in many cases increasing the MGA slack value simply pushes the model further along the same technology dimensions rather than deploying different technologies. For example, the IGCC market share in MGA Iteration 1 increases from 30% to 60% in 2050 when the slack is increased from 2% to 5% (Appendix B). Other technologies, such as concentrating solar thermal or $H_2$ fuel cell vehicles do not appear in any of the solutions. However, it is possible to modify the MGA algorithm to select certain technologies. For example, a targeted MGA run indicated that an 18% market share of solar PV was possible with 2% slack, which represents an overall cost similar to that in the 80% cap scenario. Thus, MGA can be directed by the modeler to search the decision space in a targeted fashion.

The MGA objective function weighting scheme should be adapted to the specific model context in which it is applied. In this paper, we test both an integer weighting and normalized sector method. An advantage of the latter approach is that each technology-specific MGA objective function weight is proportional to the technology's contribution in previous solutions, such that the model has the greatest incentive to deploy the technologies that play the smallest role in previous solutions. Additional weighting schemes specific to energy system models could be tested in future analysis, including grouping technologies into sets (e.g., coal, renewables) and then applying a weight to the entire group when one technology in the group is active in previous solutions. In addition, it would be worth exploring the application of different weights to each sector. For example, in the current study, we observed greater variation in electric sector deployment patterns compared to light transport, particularly at MGA slack values ranging from 1-5%. If a modeler wanted to focus on possible changes in a particular sector, different sector-specific MGA objective function weights could be applied. Many other adaptions are possible, depending on the goals of the analysis.

Observations from the MGA results can inform the approach to public policy. For example, if a $CO_2$ cap is deemed politically untenable, the results suggest that it might be plausible to achieve similar emissions reductions through higher deployment of renewables, including solar PV. Perhaps the cost to meet the 40% $CO_2$ reduction under the cap could be translated into an equivalent feed-in tariff or tax credit for renewables. Furthermore, the model results indicate that the 10% hurdle rate applied to alternative vehicle technologies may provide a drag on their deployment. Perhaps campaigns aimed at educating the public on alternative vehicle technology, including PEVs, could reduce public resistance to adoption and push the empirically-derived hurdles rates closer to zero. Accelerated PEV deployment along with clean electricity could lead to further emissions reductions. Additional scenarios reflecting reduced investment cost in renewables and a lower hurdle rate could be run to further explore the possible effect of alternative policy options. Such an iterative approach is required to properly flex the model in a way that produces useful, policy relevant insight.

Finally, we note that MGA is not a panacea for model-based uncertainty analysis. Notably, MGA results do not include probability-weighted outcomes or the value of imperfect information, as recommended by Kann and Weyant (2000). Other techniques, such as sensitivity analysis and stochastic optimization should also be utilized.

 

## 6.  Conclusions

Given the deep uncertainty associated with future energy system development, models should not be used to produce precise-looking projections that embody a high level of false precision. Rather, energy system models should be used by analysts in an iterative manner to systematically search the decision space in a way that generates insight that accounts for structural, parametric, and value-based uncertainties. MGA was developed over 30 years ago as a technique that changes the structure of mathematical models to search the model's decision space to account for unmodeled issues. Given the complexity and uncertainty associated with energy system development, MGA represents a useful way to explore the decision landscape.

The MGA results presented here highlight the false precision underlying the often limited results produced with conventional scenario analysis. Energy system models are most useful when they can be used to interactively probe the decision space in a way that challenges our mental models and leads to new insight. Such an outcome can be achieved by placing energy system models such as Temoa in a framework that allows the user to extend their own cognitive abilities by generating model results on demand. A critical element of such a framework is the capability to interactively interrogate the model by applying different methods to address uncertainty. In future work, we plan to design an interface that would allow Temoa users to apply a number of different techniques to explore energy futures, including sensitivity analysis, MGA, and stochastic optimization along with appropriate visualization of the results. In this way, energy system models could serve a more much useful role by actively engaging a wide range of users and helping them reason through different assumptions, options, and strategies. Perhaps the most useful deliverable from model-based analyses is not a set of projections, but rather a tool of exploration that allows users to interrogate the model.

## Acknowledgements

This material is based upon work supported by the National Science Foundation (CBET- 1055622).

# Appendix A:
# Case Study Input Data

 

**Figure A.1.** Network flow diagram for the modeled energy system.

**Table A.1** Technology names and descriptions used in Temoa input database

| Model Tech Name | Tech Description | Model Tech Name | Tech Description |
|---|---|---|---|
| IMPELCDSL | Imported diesel oil | T_LDV_FDSL_N | New full diesel vehicle |
| IMPELCRFL | Imported residual fuel oil | T_LDV_SSDSL_N | New small SUV diesel vehicle |
| IMPELCBSTMEA | Imported biomass | T_LDV_CDSL_HYB_N | New compact diesel hybrid vehicle |
| IMPELCMSW | Imported MSW | T_LDV_FDSL_HYB_N | New full diesel hybrid vehicle |
| IMPELCBIO | Imported biomass to bioIGCC | T_LDV_SSDSL_HYB_N | New small SUV diesel hybrid vehicle |
| IMPELCCOAB | Imported bituminous coal | T_LDV_CE85X_N | New compact E85X vehicle |
| IMPELCCOAS | Imported subbituminous coal | T_LDV_FE85X_N | New full E85X vehicle |
| IMPELCCOAL | Imported lignite coal | T_LDV_SSE85X_N | New small SUV E85X vehicle |
| IMPELCNGA | Imported natural gas | T_LDV_CE85X_HYB_N | New compact E85X hybrid vehicle |
| IMPELCURN | Imported natural uranium | T_LDV_FE85X_HYB_N | New full E85X hybrid vehicle |
| IMPTRNDSL | Imported diesel to transport | T_LDV_SSE85X_HYB_N | New small SUV E85X hybrid vehicle |
| IMPTRNE85 | Imported E85 to transport | T_LDV_CE85X_PHEV10_N | New compact E85X plugin hybrid 10 miles vehicle |
| IMPTRNE10 | Imported E10 to transport | T_LDV_FE85X_PHEV10_N | New full E85X plugin hybrid 10 miles vehicle |
| IMPTRNH2 | Imported hydrogen fuel cell to transport | T_LDV_SSE85X_PHEV10_N | New small SUV E85X plugin hybrid 10 miles vehicle |
| IMPTRNCNG | Imported CNG to transport | T_LDV_CE85X_PHEV40_N | New compact E85X plugin hybrid 40 miles vehicle |
| E_COALSTM_R | Existing coal steam power plant | T_LDV_FE85X_PHEV40_N | New full E85X plugin hybrid 40 miles vehicle |
| E_COALSTM_N | New pulverized coal steam power plant | T_LDV_SSE85X_PHEV40_N | New small SUV E85X plugin hybrid 40 miles vehicle |
| E_COALIGCC_N | New coal IGCC power plant | T_LDV_MCELC_N | New mini compact electric vehicle |
| E_COALIGCC_CCS_N | New coal IGCC with CCS power plant | T_LDV_CELC_N | New compact electric vehicle |
| E_URNLWR_R | Existing nuclear LWR power plant | T_LDV_FELC_N | New full electric vehicle |
| E_URNLWR_N | New nuclear LWR power plant | T_LDV_SSELC_N | New small SUV electric vehicle |
| E_HYDCONV_R | Existing conventional hydroelectric power plant | T_LDV_MCE10_N | New mini compact E10 vehicle |
| E_HYDREV_R | Existing reversible hydroelectric power plant | T_LDV_CE10_N | New compact E10 vehicle |
| E_OILSTM_R | Existing residual fuel oil steam power plant | T_LDV_FE10_N | New full E10 vehicle |
| E_DSLCT_R | Existing diesel combustion turbine power plant | T_LDV_SSE10_N | New small SUV E10 vehicle |
| E_DSLCC_R | Existing diesel combined cycle power plant | T_LDV_CE10_HYB_N | New compact E10 hybrid vehicle |
| E_NGASTM_R | Existing natural gas steam power plant | T_LDV_FE10_HYB_N | New full E10 hybrid vehicle |
| E_NGACT_R | Existing natural gas combustion turbine power plant | T_LDV_SSE10_HYB_N | New small SUV E10 hybrid vehicle |
| E_NGACC_R | Existing natural gas combined cycle power plant | T_LDV_CE10_PHEV10_N | New compact E10 plugin hybrid 10 miles vehicle |
| E_NGACT_N | New natural gas combustion turbine power plant | T_LDV_FE10_PHEV10_N | New full E10 plugin hybrid 10 miles vehicle |
| E_NGACC_N | New natural gas combined cycle power plant | T_LDV_SSE10_PHEV10_N | New small SUV E10 plugin hybrid 10 miles vehicle |
| E_NGAACT_N | New natural gas advanced combustion turbine power plant | T_LDV_CE10_PHEV40_N | New compact E10 plugin hybrid 40 miles vehicle |
| E_NGAACC_N | New natural gas advanced combined cycle power plant | T_LDV_FE10_PHEV40_N | New full E10 plugin hybrid 40 miles vehicle |
| E_NGACC_CCS_N | New natural gas combined cycle with CCS power plant | T_LDV_SSE10_PHEV40_N | New small SUV E10 plugin hybrid 40 miles vehicle |
| E_NGA_ALL_N | An aggregation of all new NGA-based plants | T_LDV_CH2FC_N | New compact hydrogen fuel cell vehicle |
| E_BIOSTM_R | Existing biomass steam power plant | T_LDV_FH2FC_N | New full hydrogen fuel cell vehicle |
| E_BIOIGCC_N | New bioIGCC power plant | T_LDV_SSH2FC_N | New small SUV hydrogen fuel cell vehicle |
| E_GEO_R | Existing geothermal power plant | T_LDV_CCNG_N | New compact CNG vehicle |
| E_GEOBCFS_N | New geothermal binary cycle & flashed system power plant | T_LDV_FCNG_N | New full CNG vehicle |
| E_MSWSTM_R | Existing municipal solid waste steam power plant | T_LDV_MCE10_R | Existing mini compact E10 vehicle |
| E_WND_R | Existing wind power plant | T_LDV_CE10_R | Existing compact E10 vehicle |
| E_WNDCL4_N | New wind class 4 power plant | T_LDV_FE10_R | Existing full E10 vehicle |
| E_WNDCL5_N | New wind class 5 power plant | T_LDV_SSE10_R | Existing small SUV E10 vehicle |
| E_WNDCL6_N | New wind class 6 power plant | T_LDV_FDSL_R | Existing full diesel vehicle |
| E_SOLPV_R | Existing solar photovoltaic power plant | T_LDV_FE85X_R | Existing full E85X vehicle |
| E_SOLTH_R | Existing solar thermal power plant | T_LDV_FELC_R | Existing full electric vehicle |
| E_SOLPVCEN_N | New solar photovoltaic centralized power plant | T_LDV_FCNG_R | Existing full CNG vehicle |
| E_SOLTHCEN_N | New solar thermal centralized power plant | T_EA_DSL | CO2 emission accounting tech for diesel |
| E_EA_COAB | CO2 emission accounting tech for coal bituminous | T_EA_E85X | CO2 emission accounting tech for E85 |
| E_EA_COAS | CO2 emission accounting tech for coal subbituminous | T_EA_E10 | CO2 emission accounting tech for E10 |
| E_EA_COAL | CO2 emission accounting tech for coal lignite | T_EA_CNG | CO2 emission accounting tech for CNG |
| E_CCR_COALSTM_N | CO2 capture retrofit tech before new coal steam plant | T_BLND_E10_PHEV10 | Blending tech to collect E10 and ELC for PHEV10 |
| E_CCR_COALIGCC_N | CO2 capture retrofit tech before coal IGCC plant | T_BLND_E10_PHEV40 | Blending tech to collect E10 and ELC for PHEV40 |
| E_BLND_BITSUBLIG | Blending tech to collect bit sub lig | T_BLND_E85X_PHEV10 | Blending tech to collect E85 and ELC for PHEV10 |

| O_ELCDEM | Electricity demand technology for other sectros than transport | T_BLND_E85X_PHEV40 | Blending tech to collect E85 and ELC for PHEV40 |
| T_LDV_CDSL_N | New compact diesel vehicle | T_LDV_BLNDDEM | Blending LDV demand tech for MC-C-F-SS size classes |

**Table A.2.** Existing capacity (in GW) and lifetime of electric power plants

| Technology Name | Lifetime (years) | 2015 | 2020 | 2025 | 2030 | 2035 | 2040 | 2045 | 2050 |
|---|---|---|---|---|---|---|---|---|---|
| Oil Steam (Residual Fuel Oil LS), Existing | 50 | 15.51 | 15.51 | 15.51 | 15.51 | 15.51 | 15.51 | 15.51 | 15.51 |
| Natural Gas Steam, Existing | 50 | 78.09 | 78.09 | 78.09 | 78.09 | 78.09 | 78.09 | 78.09 | 78.09 |
| Diesel Oil Combustion Turbine, Existing | 30 | 26.37 | 26.37 | 26.37 | 26.37 | 26.37 | | | |
| Natural Gas Combustion Turbine, Existing | 30 | 112.23 | 112.23 | 112.23 | 112.23 | 112.23 | | | |
| Diesel Oil Combined-Cycle, Existing | 30 | 6.83 | 6.83 | 6.83 | 6.83 | 6.83 | | | |
| Natural Gas Combined-Cycle, Existing | 30 | 191.87 | 191.87 | 191.87 | 191.87 | 191.87 | | | |
| Wood/Biomass Steam, Existing | 25 | 3.45 | 3.45 | 3.45 | 3.45 | | | | |
| Municipal Solid Waste Steam, Existing | 30 | 3.80 | 3.80 | 3.80 | 3.80 | 3.80 | | | |
| Geothermal, Existing | 25 | 2.65 | 2.65 | 2.65 | 2.65 | | | | |
| Hydroelectric, Conventional, Existing | 50 | 78.64 | 78.64 | 78.64 | 78.64 | 78.64 | 78.64 | 78.64 | 78.64 |
| Hydroelectric, Reversible, Existing | 50 | 22.40 | 22.40 | 22.40 | 22.40 | 22.40 | 22.40 | 22.40 | 22.40 |
| Wind, Existing | 25 | 75.80 | 75.80 | 75.80 | 75.80 | | | | |
| Solar Thermal, Existing | 30 | 1.77 | 1.77 | 1.77 | 1.77 | 1.77 | | | |
| Solar Photovoltaic, Existing | 30 | 10.41 | 10.41 | 10.41 | 10.41 | 10.41 | | | |
| Residual Coal Steam, Existing | 50 | 281.40 | 281.40 | 281.40 | 281.40 | 281.40 | 281.40 | 281.40 | 281.40 |
| Pre-Existing Nuclear LWRs | 50 | 99.60 | 99.60 | 99.60 | 99.60 | 99.60 | 99.60 | 99.60 | 99.60 |
| Nuclear LWRs | 50 | | | | | | | | |
| Integrated Coal Gasification Combined Cycle CO2 Capture | 50 | | | | | | | | |
| Natural Gas Combined Cycle CO2 Capture | 30 | | | | | | | | |
| Solar PV Centralized Generation | 30 | | | | | | | | |
| Solar Thermal Centralized Generation | 30 | | | | | | | | |
| Wind Generation Class 4 | 25 | | | | | | | | |
| Wind Generation Class 5 | 25 | | | | | | | | |
| Wind Generation Class 6 | 25 | | | | | | | | |
| Natural Gas - Advanced Combined-Cycle (Turbine) | 30 | | | | | | | | |
| Natural Gas - Advanced Combustion Turbine | 30 | | | | | | | | |
| Geothermal - Binary Cycle and Flashed Steam | 25 | | | | | | | | |
| Biomass Integrated Gasification Combined-Cycle | 35 | | | | | | | | |
| Pulverized Coal Steam | 50 | | | | | | | | |
| Integrated Coal Gasification Combined Cycle | 50 | | | | | | | | |
| Natural Gas - Combined Cycle (Turbine) | 30 | | | | | | | | |
| Natural Gas - Combustion Turbine | 30 | | | | | | | | |

**Table A.3.** Upper bound values on electricity generation capacity from new geothermal and wind (GW)

| Technology Name | 2020 | 2025 | 2030 | 2035 | 2040 | 2045 | 2050 |
|---|---|---|---|---|---|---|---|
| Geothermal - Binary Cycle and Flashed Steam | 6.05 | 7.84 | 9.63 | 11.43 | 13.22 | 15.00 | 16.80 |
| Wind Generation Class 4 | 2562 | 2562 | 2562 | 2562 | 2562 | 2562 | 2562 |
| Wind Generation Class 5 | 468 | 468 | 468 | 468 | 468 | 468 | 468 |
| Wind Generation Class 6 | 108 | 108 | 108 | 108 | 108 | 108 | 108 |

The upper bound on geothermal capacity is based on AEO2012 and wind is drawn from Shay et al. (2006)

**Table A.4.** Upper bound on electricity generation capacity growth rate (%) and seed (GW)

| Technology Name | Seed[1] (GW) | Annual Growth rate (%) |
|---|---|---|
| Nuclear LWRs | 5 | 9.5% |
| Integrated Coal Gasification Combined Cycle $CO_2$ Capture | 30 | 10% |
| Natural Gas (Total Simple- and Combined Cycle) | 30 | 10% |
| Solar PV Centralized Generation | 30 | 10% |
| Solar Thermal Centralized Generation | 30 | 10% |
| Wind Generation Class 4 | 10 | 10% |
| Wind Generation Class 5 | 10 | 10% |
| Wind Generation Class 6 | 10 | 10% |
| Geothermal - Binary Cycle and Flashed Steam | 30 | 10% |
| Biomass Integrated Gasification Combined-Cycle | 30 | 10% |
| Integrated Coal Gasification Combined Cycle | 30 | 10% |
| Pulverized Coal Steam | 1 | 10% |

[1] The maximum installed capacity that the model starts to build in any time period.

**Table A.5.** Electricity demand in the non-transportation related end-use sectors (PJ)

| Commodity Description | 2015 | 2020 | 2025 | 2030 | 2035 | 2040 | 2045 | 2050 |
|---|---|---|---|---|---|---|---|---|
| Electricity demand for non-LDVs | 14410 | 15143 | 15651 | 16147 | 16576 | 17140 | 17746 | 18372 |

**Table A.6.** Existing capacity of light duty vehicles in billion vehicle miles (bnvmt)

| Technology Name | 2015 | 2020 | 2025 |
|---|---|---|---|
| Existing Mini compact conventional gasoline | 28.6 | 14.3 | 0.0 |
| Existing Compact conventional gasoline | 485.4 | 242.7 | 0.0 |
| Existing Full Diesel | 3.4 | 1.7 | 0.0 |
| Existing Full conventional gasoline | 402.6 | 201.3 | 0.0 |
| Existing Small SUV conventional gasoline | 124.6 | 62.3 | 0.0 |
| Existing Full Ethanol Flex Fuel | 79.4 | 39.7 | 0.0 |
| Existing Full CNG | 2.2 | 1.1 | 0.0 |
| Existing Full Electric | 0.2 | 0.1 | 0.0 |

**Table A.7.** Demand values for light duty transportation sector (bnvmt)

| Commodity Description | 2015 | 2020 | 2025 | 2030 | 2035 | 2040 | 2045 | 2050 |
|---|---|---|---|---|---|---|---|---|
| Total miles demanded for 4 LDV size classes | 1689 | 1799 | 1916 | 2041 | 2174 | 2315 | 2466 | 2626 |

**Table A.8.** The fixed share of LDV size classes (%)

| Time Period (t) | Mini-Compact | Compact | Full | Small SUV |
|---|---|---|---|---|

| 2015-2050 | 3% | 32% | 45% | 20% |
|-----------|-----|------|------|------|

**Table A.9.** The upper bound values on electricity generation from existing coal power plants (PJ)

| Technology Name | 2015 | 2020 | 2025 | 2030 | 2035 | 2040 | 2045 | 2050 |
|-----------------|------|------|------|------|------|------|------|------|
| Existing coal-fired steam | 5602 | 6012 | 6066 | 6026 | 5994 | 5987 | 5987 | 5987 |

**Table A.10.** The upper bound values on corn-based ethanol imports (PJ)

| Technology Name | 2015 | 2020 | 2025 | 2030 | 2035 | 2040 | 2045 | 2050 |
|-----------------|------|------|------|------|------|------|------|------|
| Ethanol import | 1751 | 2604 | 3143 | 5024 | 5024 | 5024 | 5024 | 5024 |

# Appendix B:
# Complete Results from MGA Integer and Normalized Sector Weighting Methods

**MGA: Integer Weighting with 1% Slack**

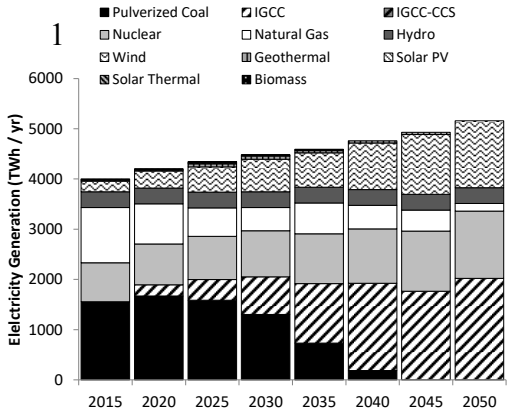

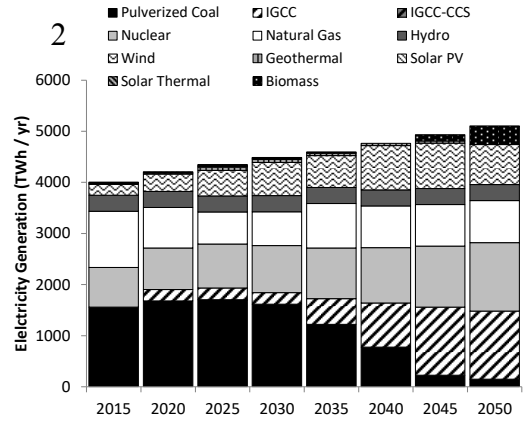

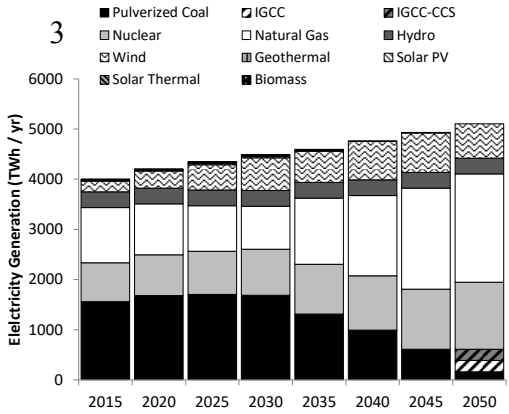

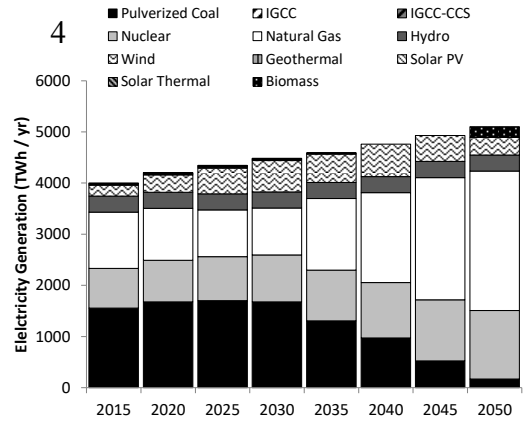

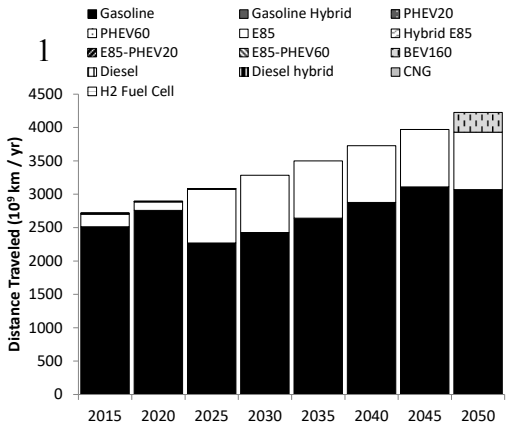

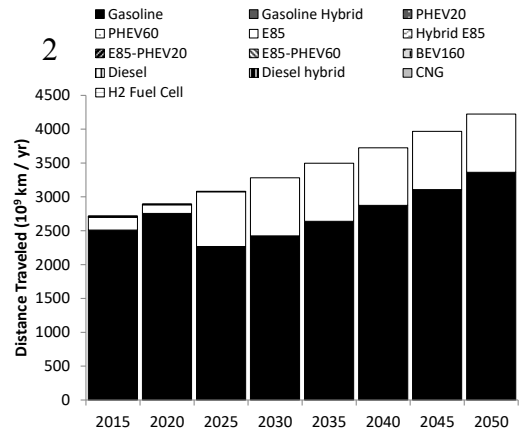

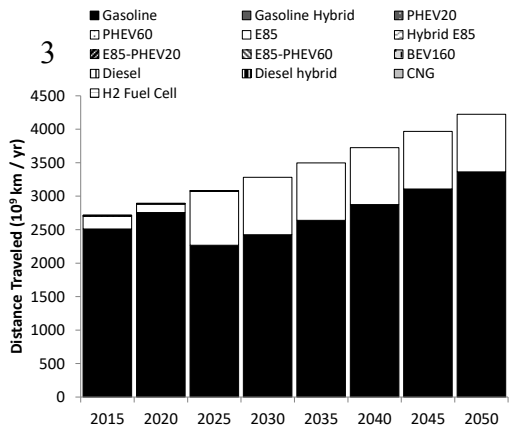

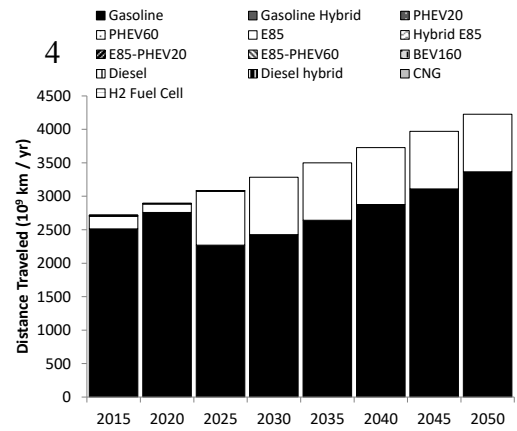

**MGA: Integer Weighting with 2% Slack**

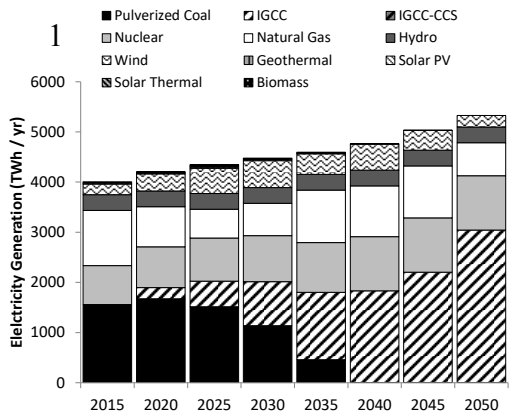

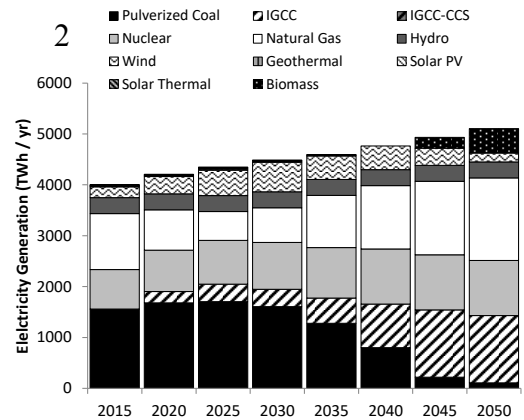

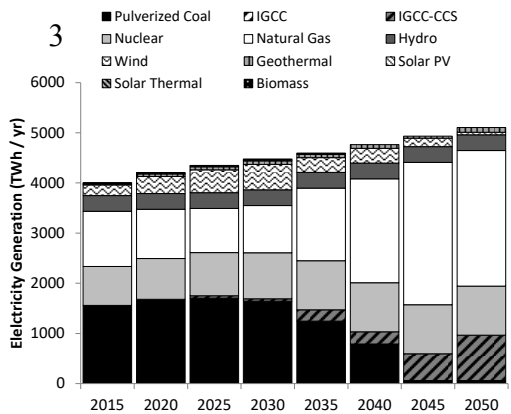

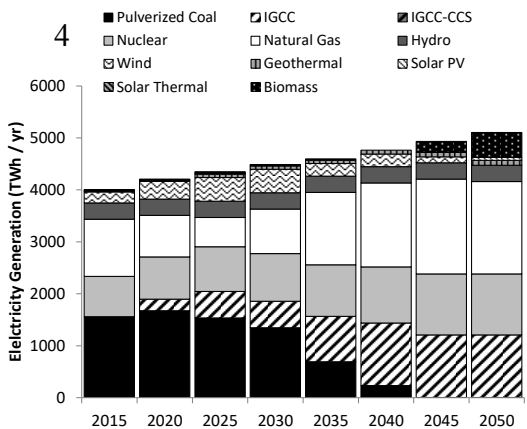

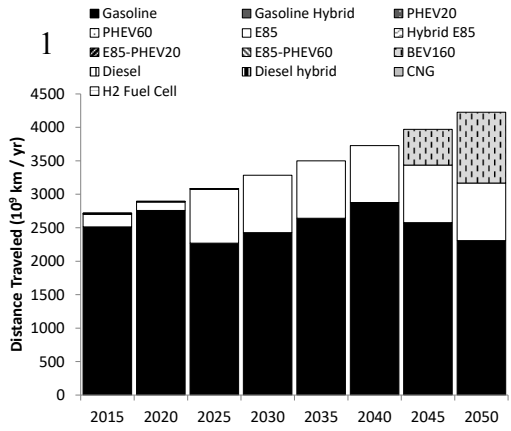

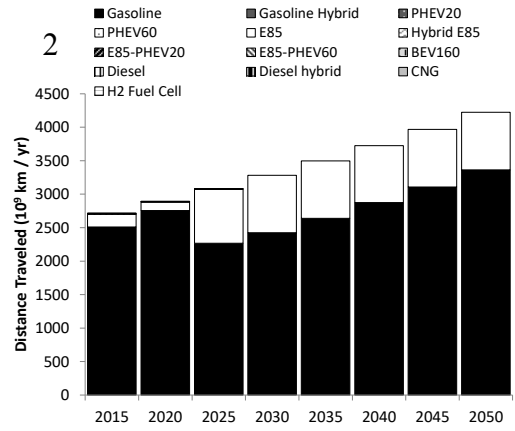

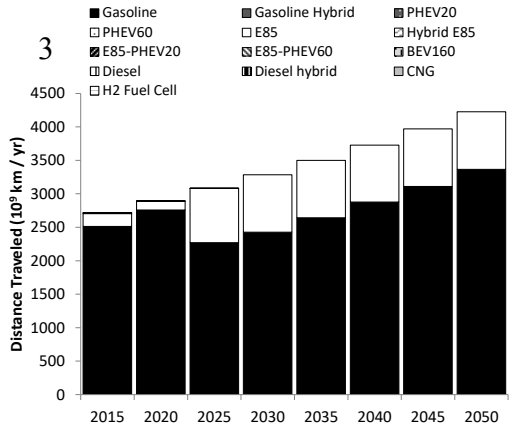

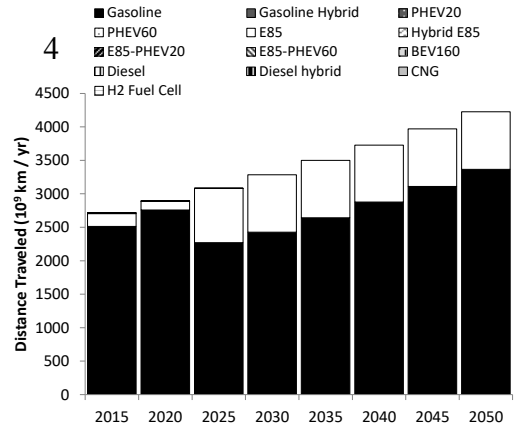

**MGA: Integer Weighting with 5% Slack**

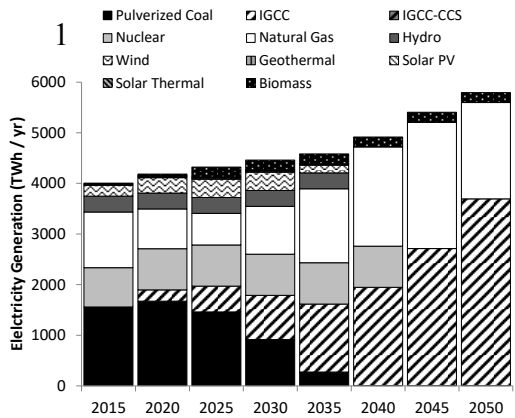

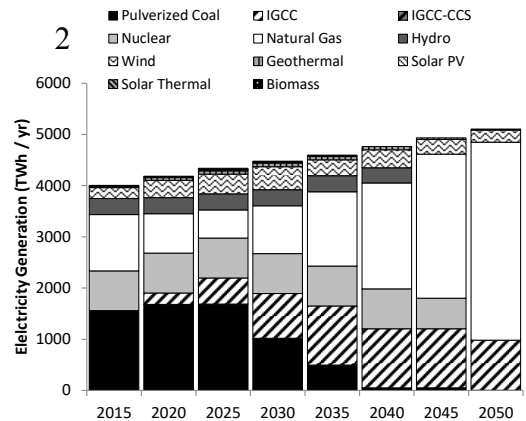

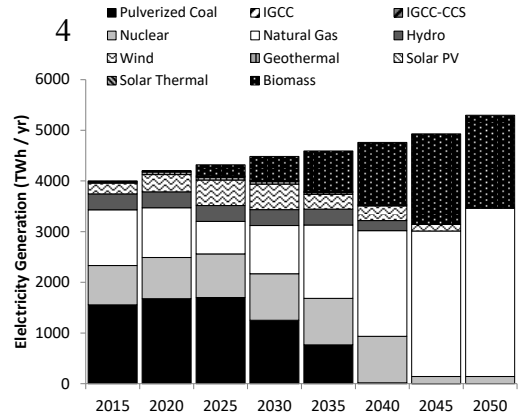

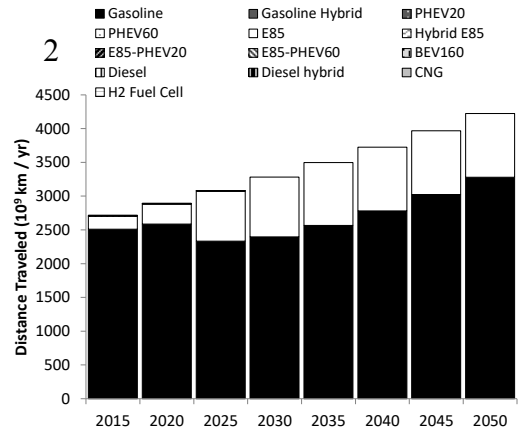

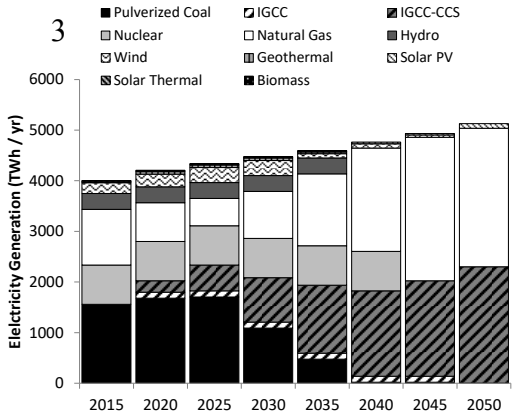

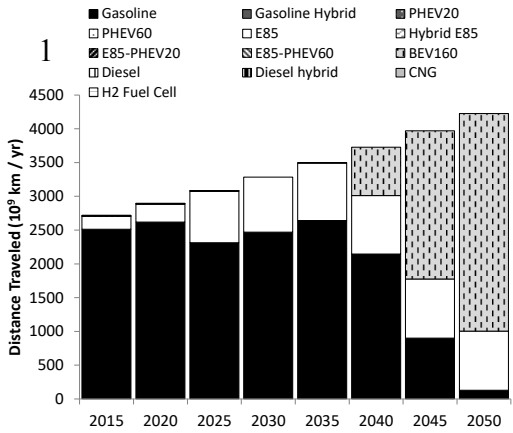

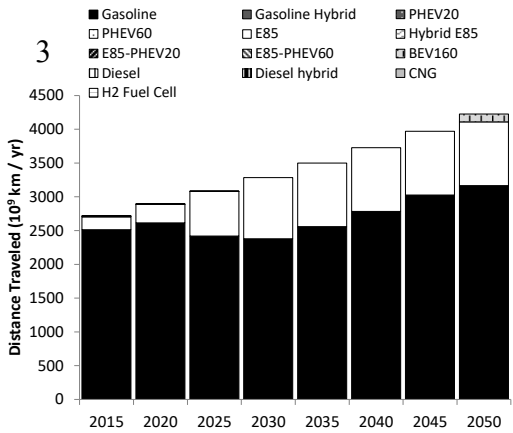

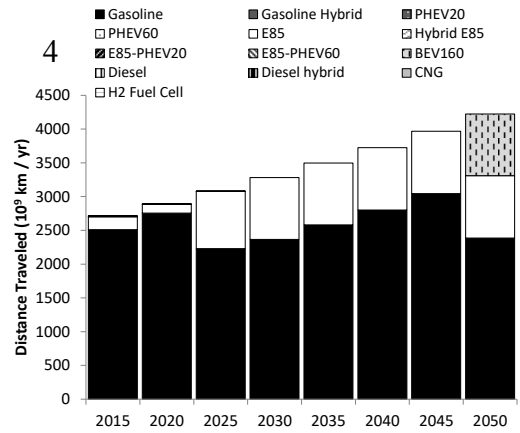

**MGA: Integer Weighting with 10% Slack**

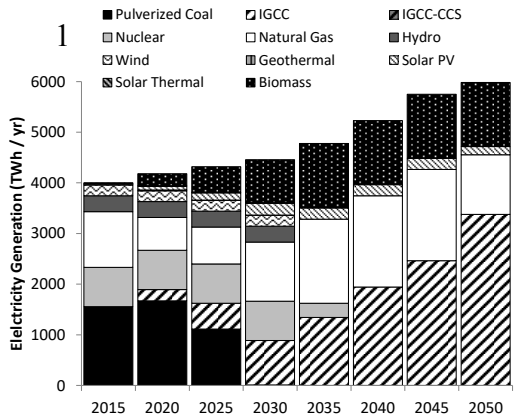

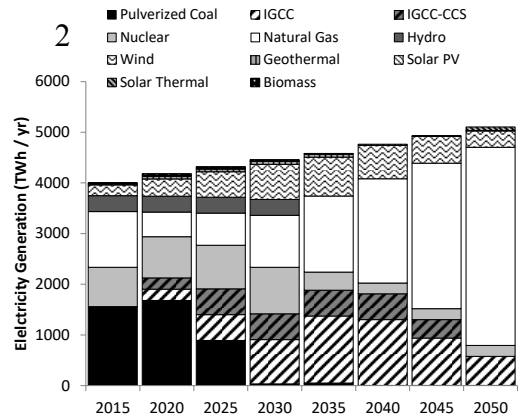

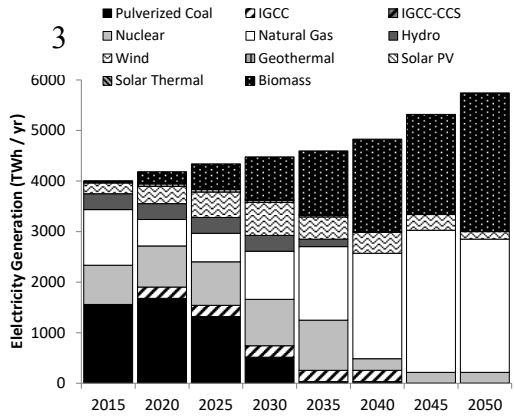

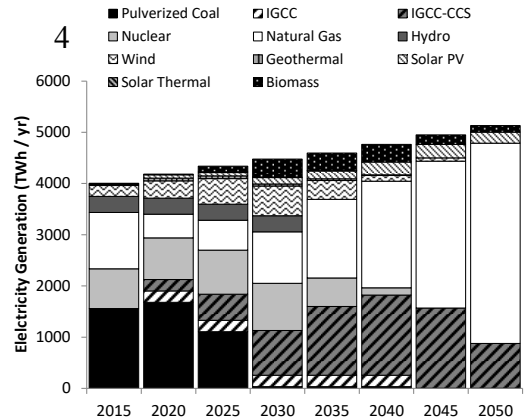

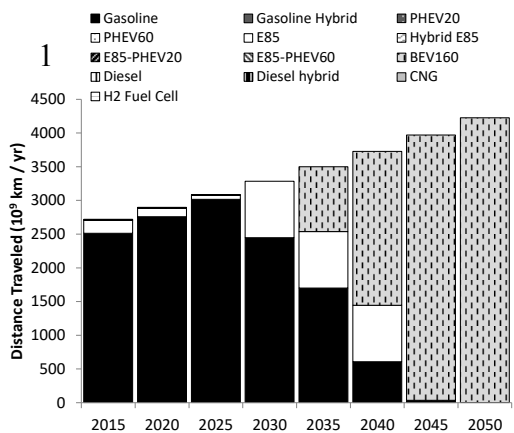

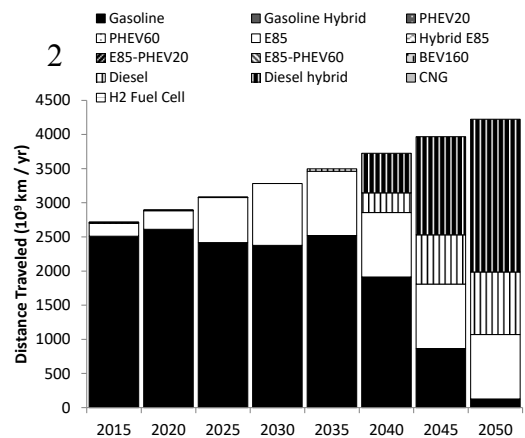

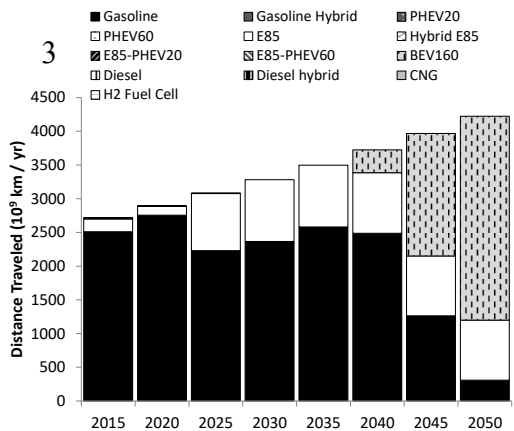

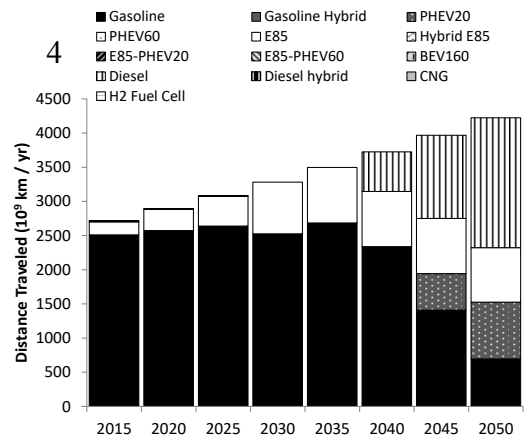

 

**MGA: Normalized Weighting by Sector with 1% Slack**

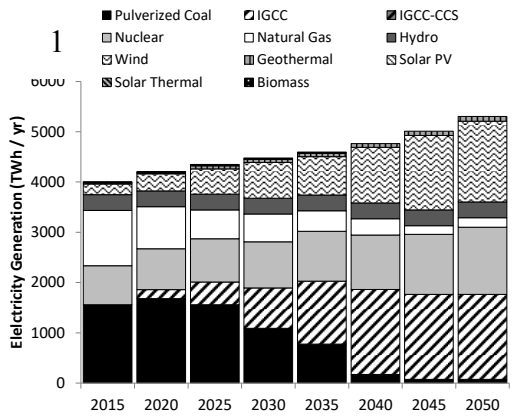

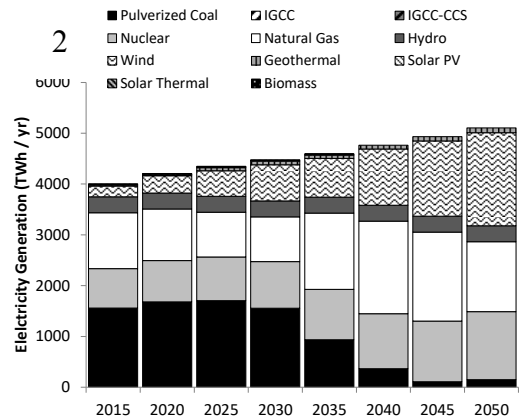

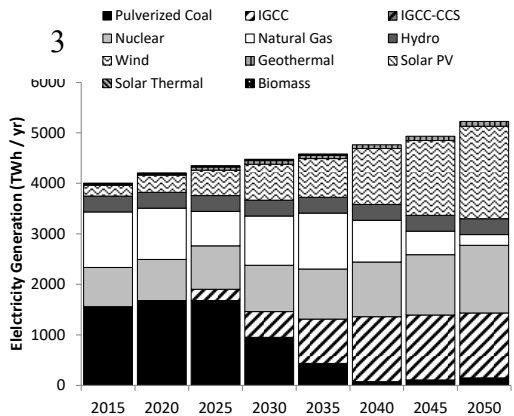

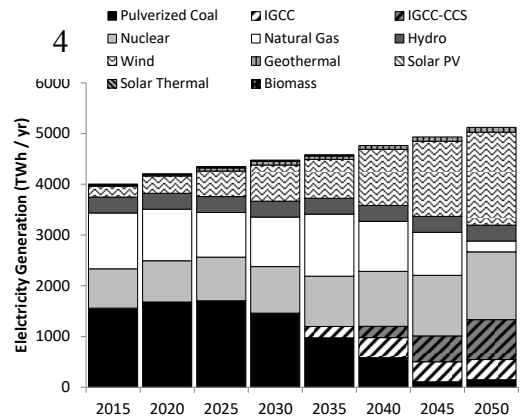

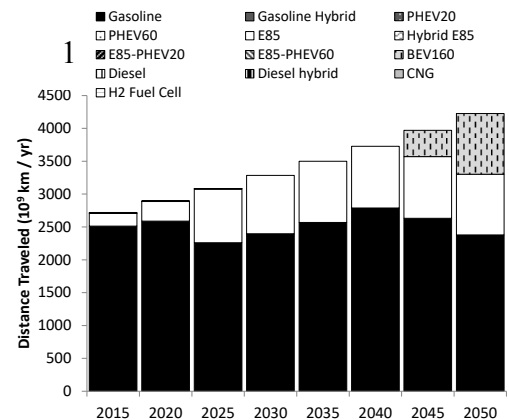

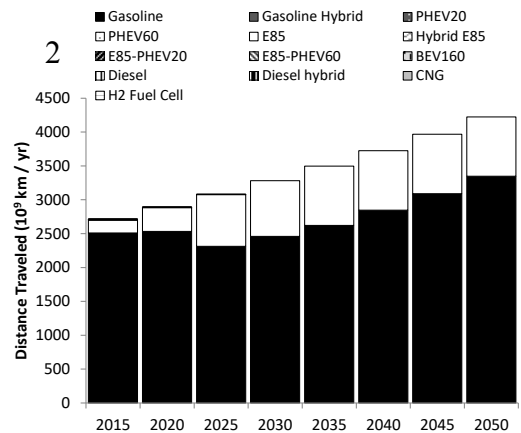

 

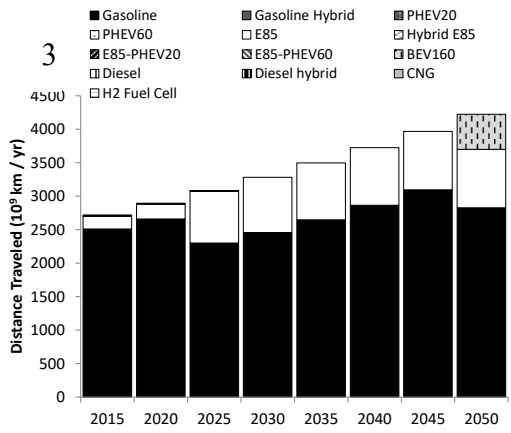

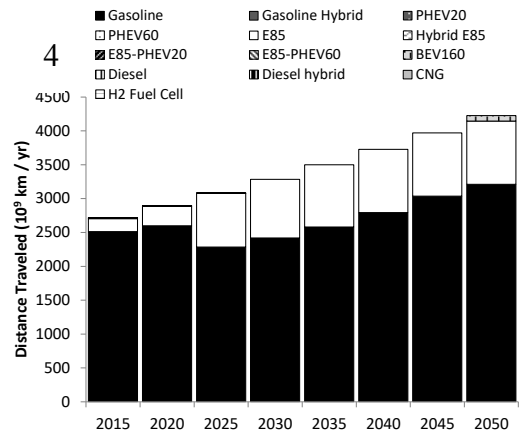

## MGA: Normalized Weighting by Sector with 2% Slack

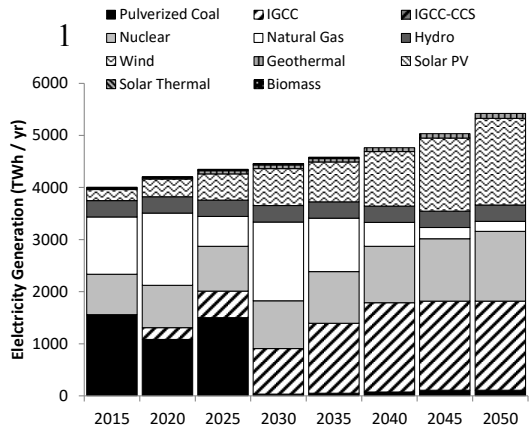

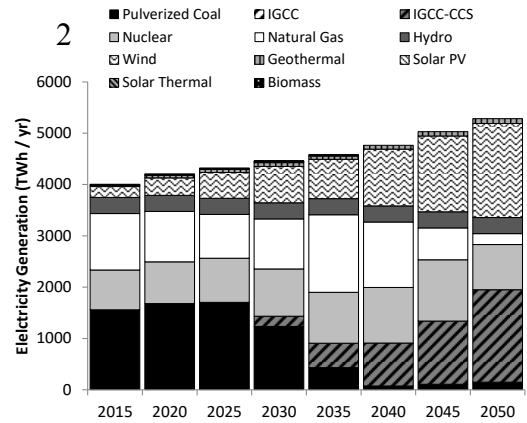

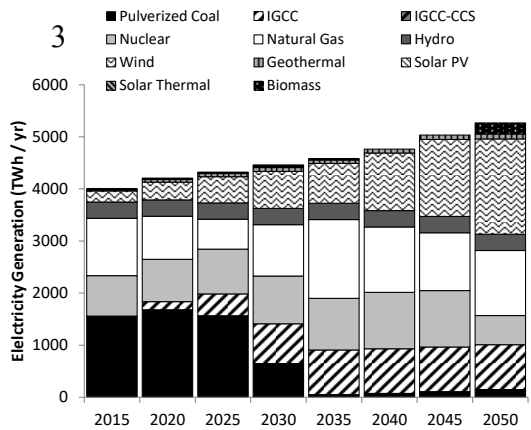

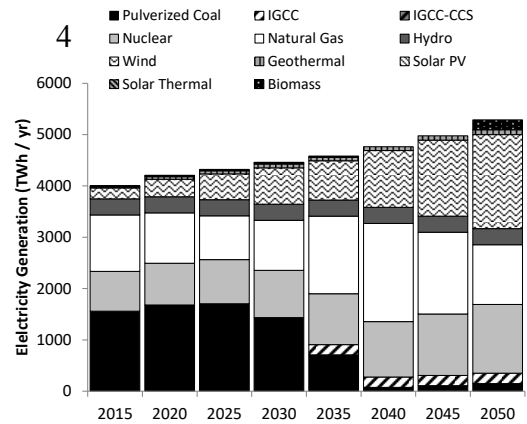

 

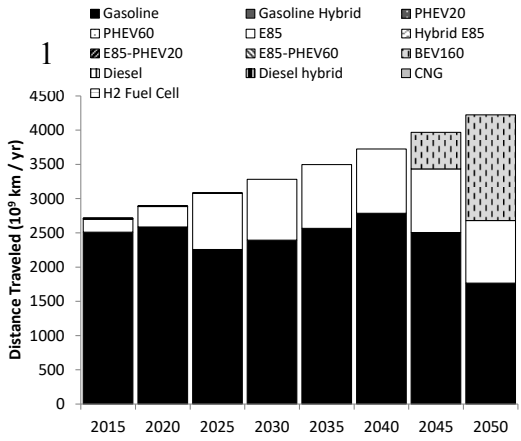

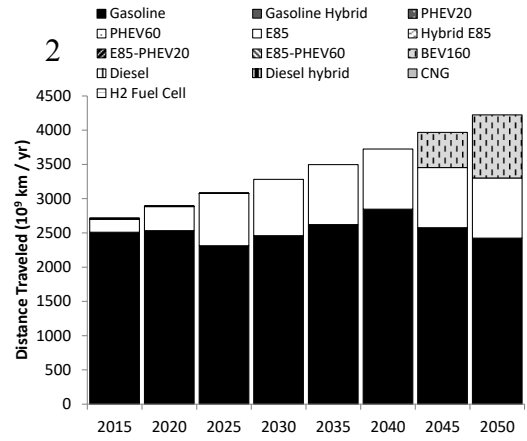

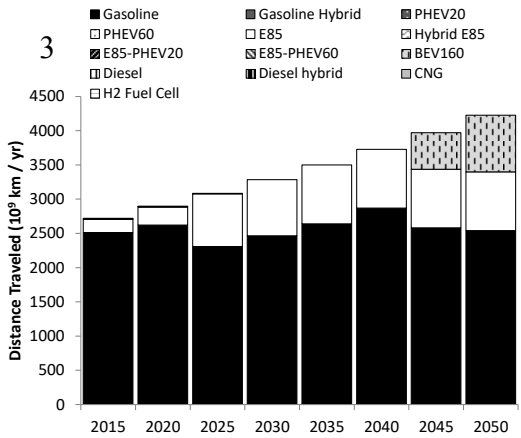

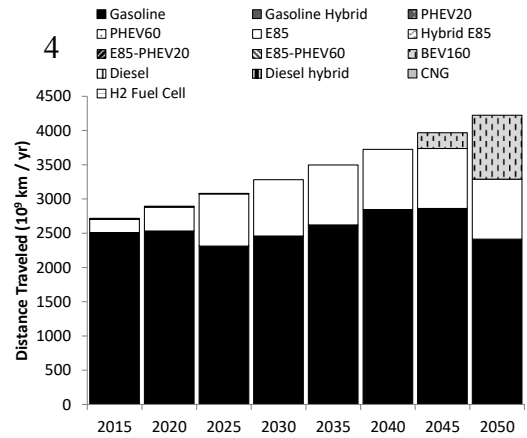

## MGA: Normalized Weighting by Sector with 5% Slack

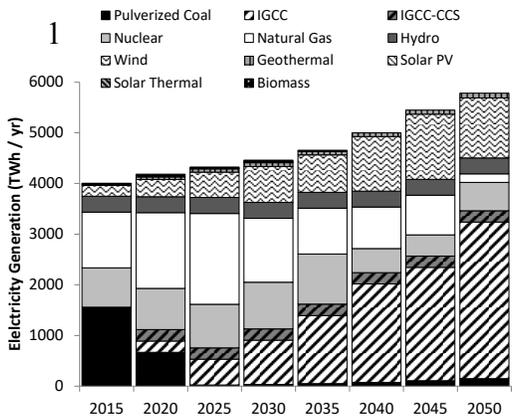

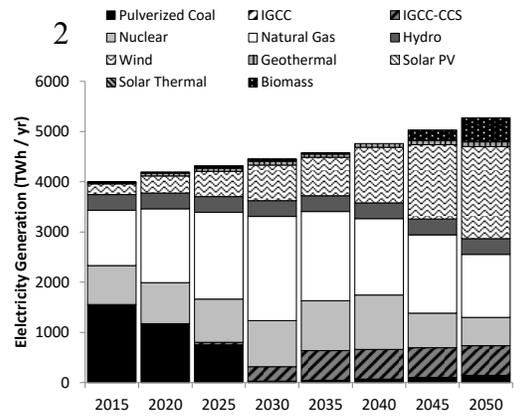

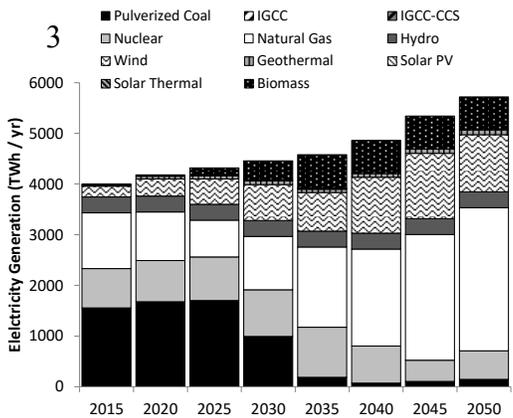

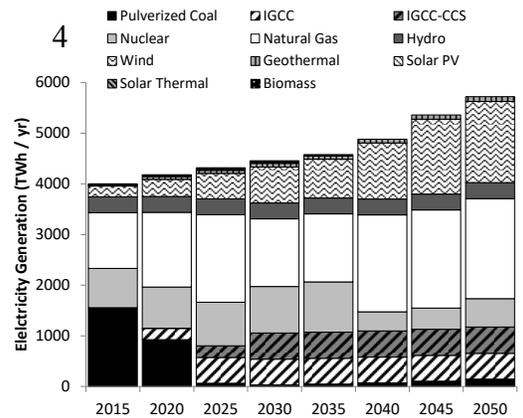

 

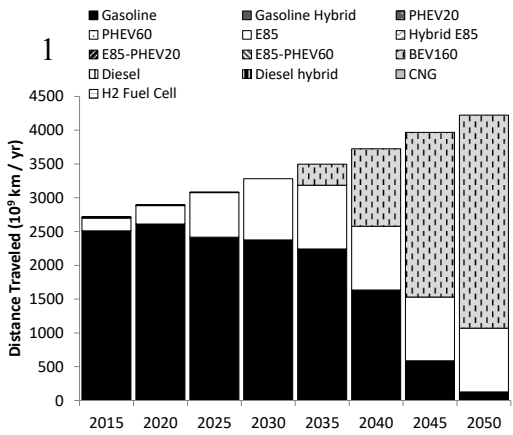

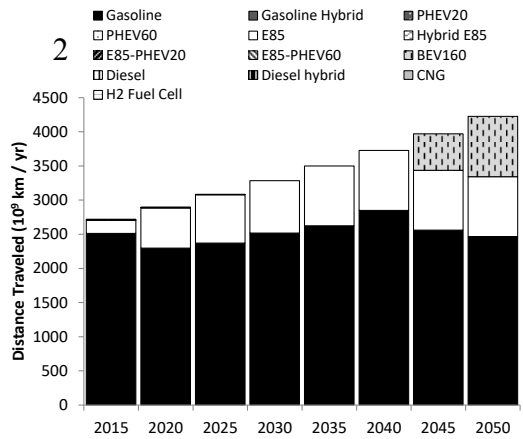

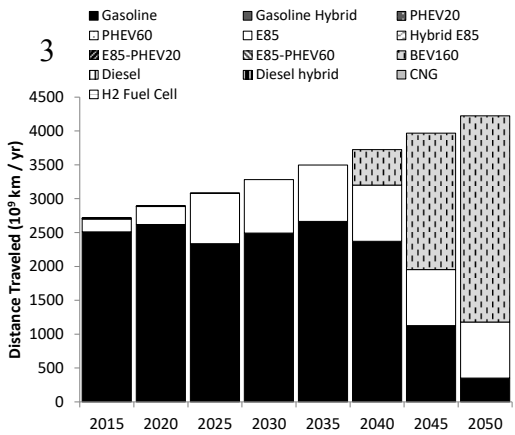

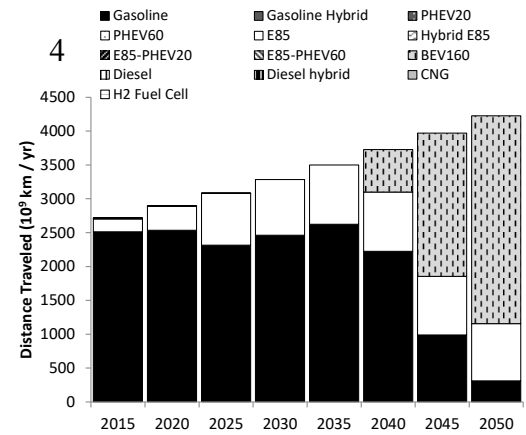

**MGA: Normalized Weighting by Sector with 10% Slack**

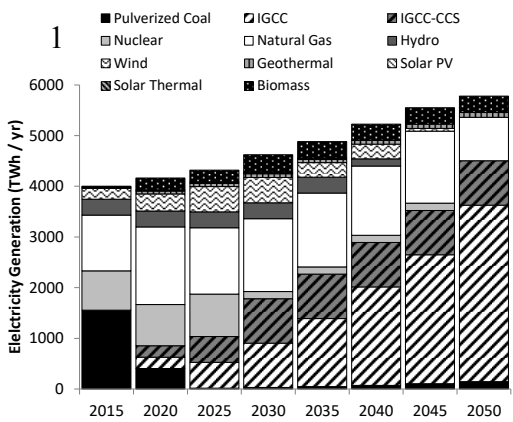

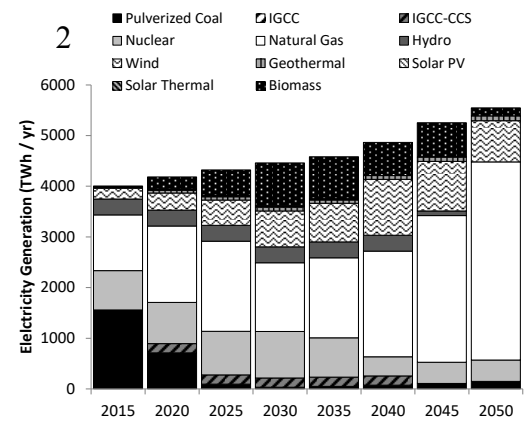

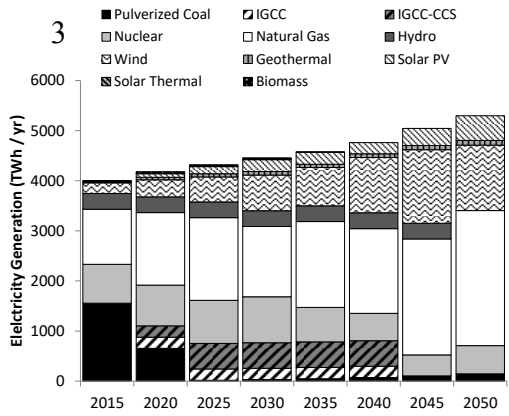

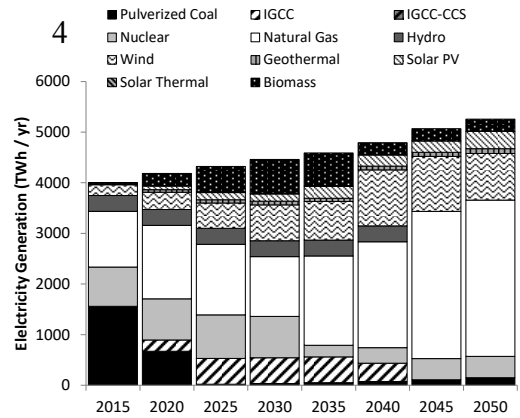

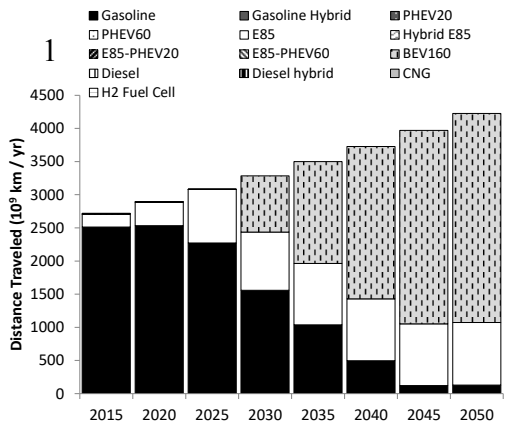

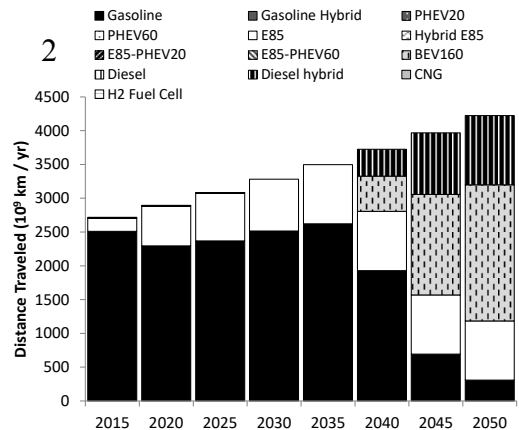

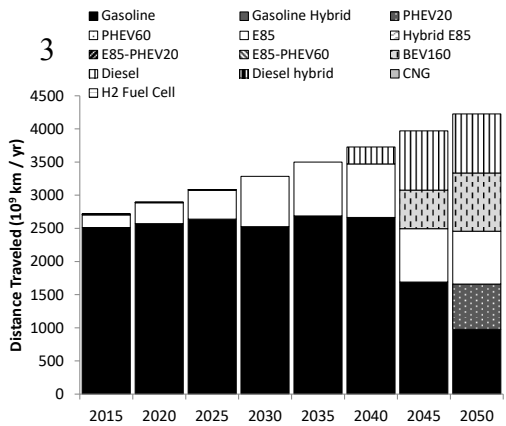

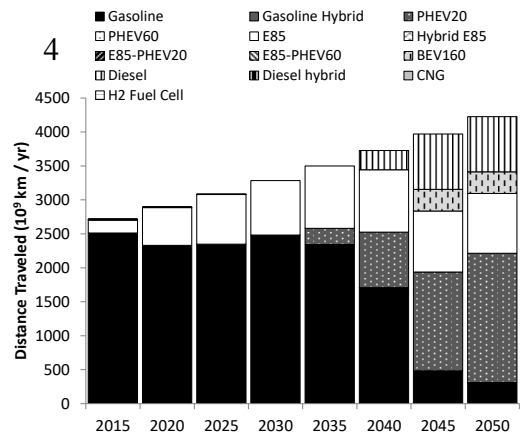

# Appendix C:
# Results From Discount Rate Tests

## Moderate CO₂ Cap: 0.1% Discount Rate

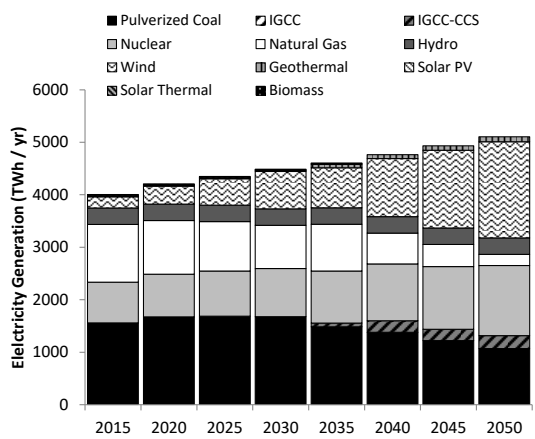

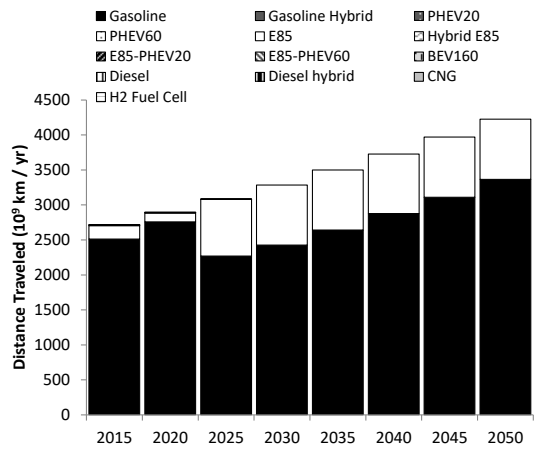

## Moderate CO₂ Cap: 10% Discount Rate

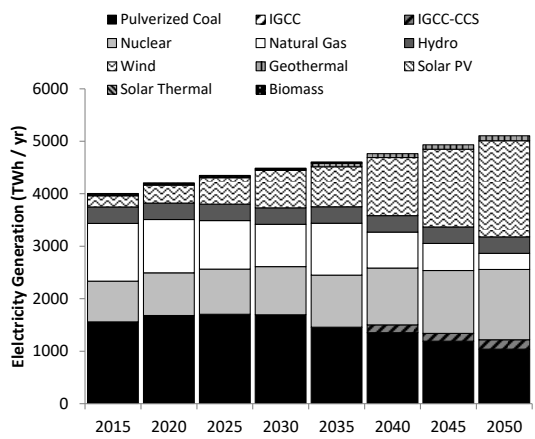

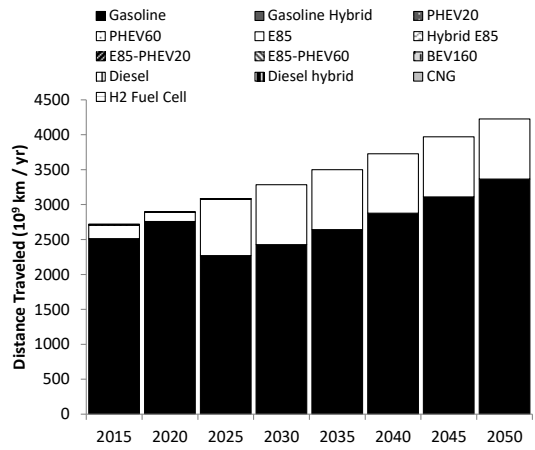

# MGA: Normalized Weighting by Sector − 2% Slack − 0.1% Discount Rate

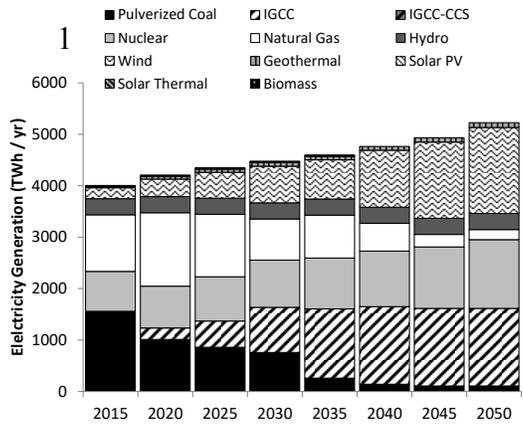

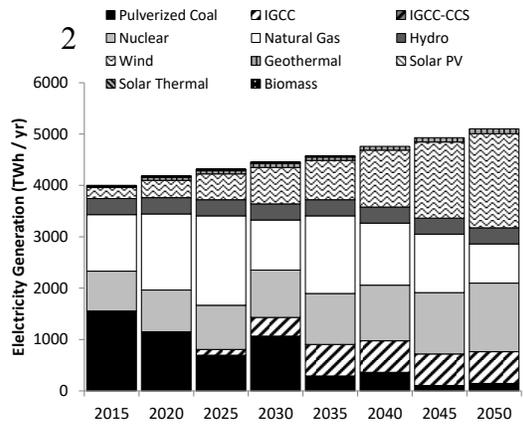

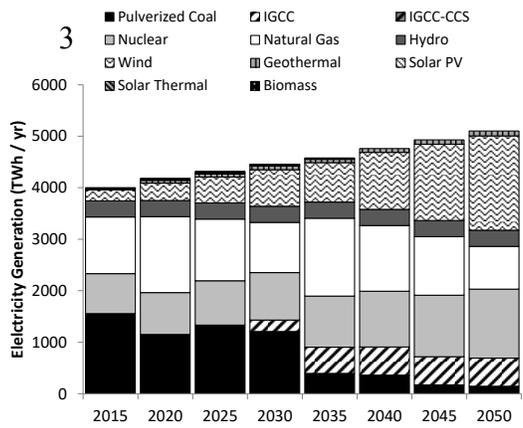

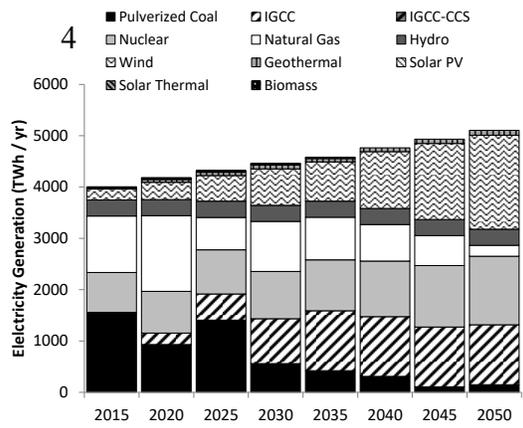

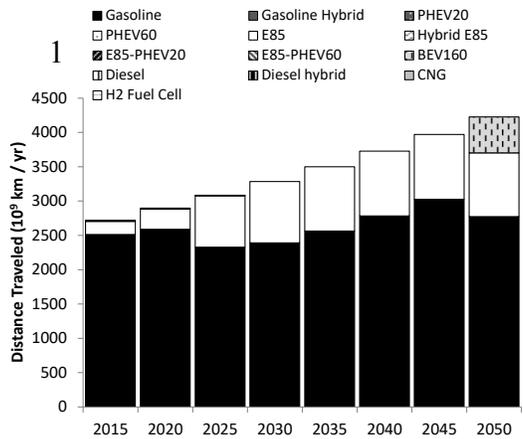

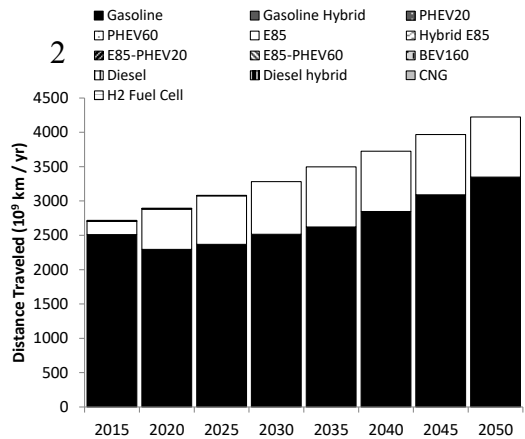

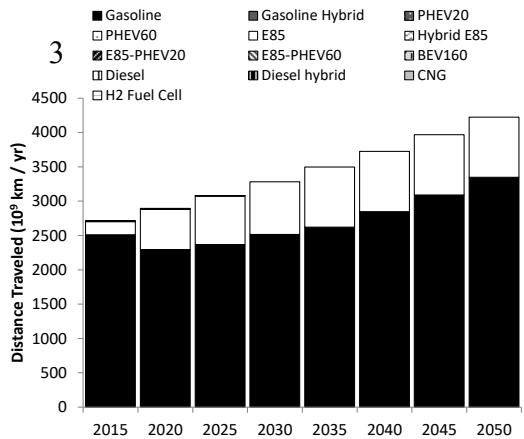

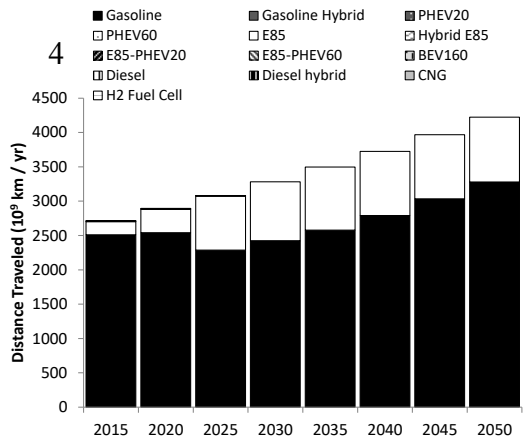

## MGA: Normalized Weighting by Sector – 2% Slack – 10% Discount Rate

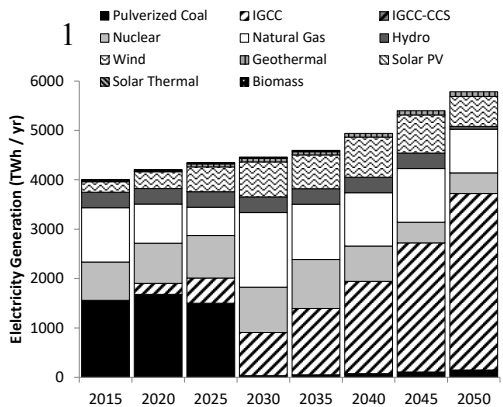

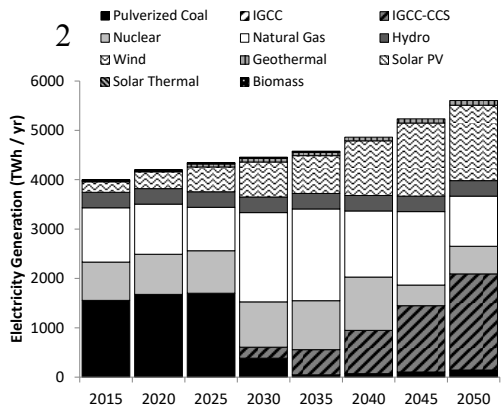

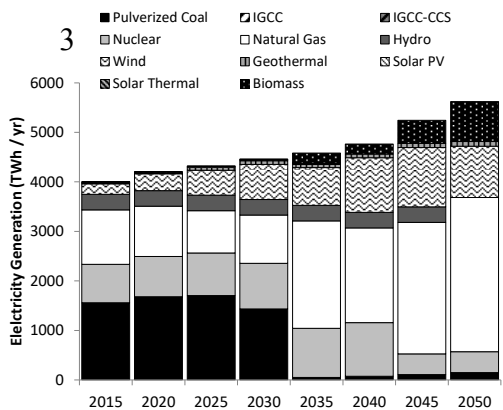

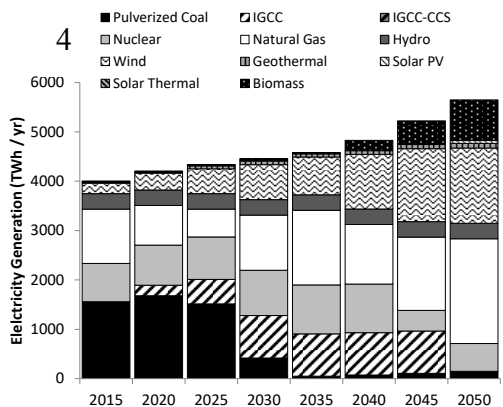

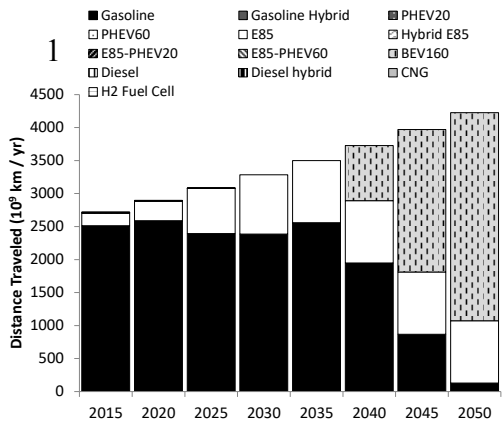

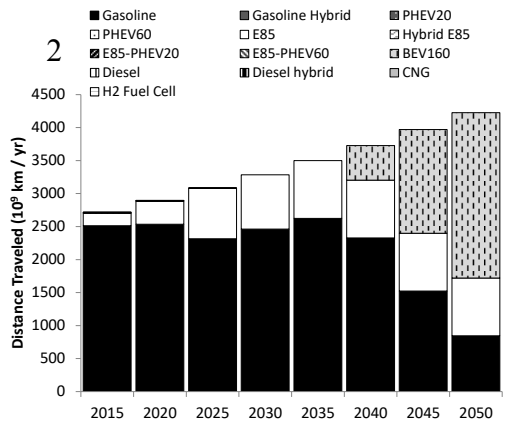

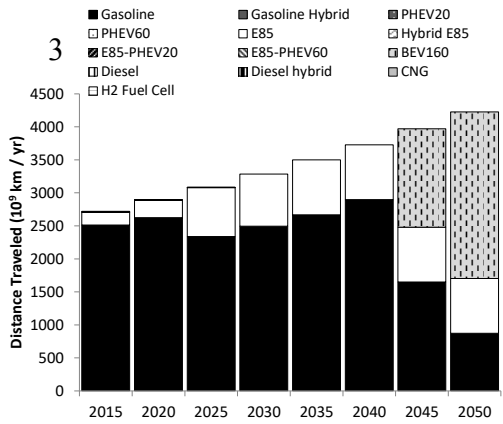

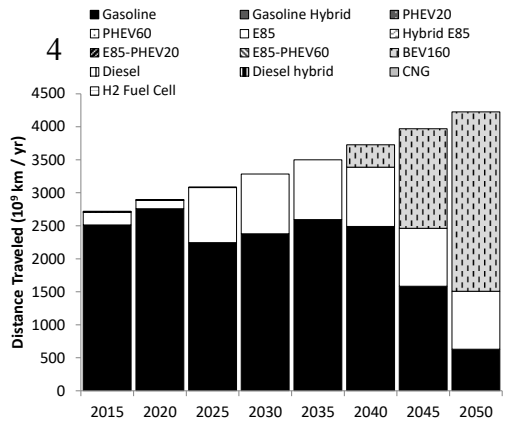